\documentclass[10pt,a4paper]{article}

\usepackage{amsmath}

\usepackage{color}     
\usepackage{multicol}  
\usepackage{anysize}
\usepackage{fancyhdr}
\usepackage{wasysym}
\usepackage{type1cm}
\usepackage{graphicx}
\usepackage{setspace}
\usepackage[font={small,singlespacing}, bf,centerlast]{caption}

\usepackage[linesnumbered]{algorithm2e}
\usepackage[numbers]{natbib}

\usepackage{xr}

\marginsize{1.5cm}{1.5cm}{0.5cm}{2cm}

\definecolor{blue4}{rgb}{0,0,0.5}
\definecolor{indianred3}{rgb}{0.8,0.33,0.33}
\definecolor{gainsboro}{rgb}{0.86,0.86,0.86}
\definecolor{dodgerblue4}{rgb}{0.06,0.31,0.55}

\makeatletter

\newenvironment{figurehere}
{\def\@captype{figure}}
{}
\makeatother

\begin{document}

\begin{center}

{\Large  Inferring repeat protein energetics from evolutionary information}\\
\vspace*{.5pt}
\end{center}
\medskip

\noindent\textbf{Espada Roc\'io$^{12}$,Parra R. Gonzalo$^{3}$,Mora Thierry$^{4}$,Walczak Aleksandra M.$^{5}$, Ferreiro Diego U.$^{12*}$ }\\
{\small
\singlespace
\noindent $^{1}$ Universidad de Buenos Aires, Facultad de Ciencias Exactas y Naturales, Departamento de Química Biológica. Buenos Aires, Argentina.\\
\noindent $^{2}$ CONICET - Universidad de Buenos Aires. Instituto de Química Biológica de la Facultad de Ciencias Exactas y Naturales (IQUIBICEN). Buenos Aires, Argentina.\\
\noindent $^{3}$ Quantitative and Computational Biology Group, Max Planck Institute for Biophysical Chemistry, Goettingen, Germany\\
\noindent $^{4}$ Laboratoire de physique statistique, Ecole Normale Sup\'erieure, CNRS and UPMC, 75005 Paris, France\\
\noindent $^{5}$ CNRS and Laboratoire de Physique Th\'eorique, Ecole Normale Sup\'erieure, Paris, France\\
\noindent $^{*}$ Corresponding author Ferreiro Diego U., mail: ferreiro@qb.fcen.uba.ar.

February 2017
}

%
%
%

\hspace{10pt}

\textbf{Abstract}

Natural protein sequences contain a record of their history. A common constraint in a given protein family is the ability to fold to specific structures, and it has been shown possible to infer the main native ensemble by analyzing covariations in extant sequences. Still, many natural proteins that fold into the same structural topology show different stabilization energies, and these are often related to their physiological behavior. We propose a description for the energetic variation given by sequence modifications in repeat proteins, systems for which the overall problem is simplified by their inherent symmetry. We explicitly account for single amino acid and pair-wise interactions and treat higher order correlations with a single term. We show that the resulting evolutionary field can be interpreted with structural detail. We trace the variations in the energetic scores of natural proteins and relate them to their experimental characterization. 
The resulting energetic evolutionary field allows the  prediction of the folding free energy change for several mutants, and can be used to generate synthetic sequences that are statistically indistinguishable from the natural counterparts.

\textbf{Keywords:} repeat proteins, coevolution analysis, ankyrin repeat, leucine-rich repeat, tetratricopeptide repeat, $\Delta$G prediction.

\hspace{10pt}

\begin{multicols}{2}
\section*{Introduction}

\begin{figure*}[!hb]
\centering
\includegraphics[width=.9\textwidth]{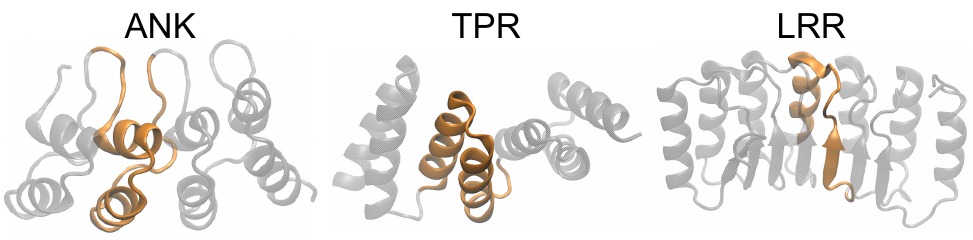}
\caption{Repeat proteins are elongated objects with internal symmetry. Representative structures of members of the repeat proteins families studied. On left, ankyrin repeat (PDB id:1N0R \citep{peng2002pnas}), center tetratricopeptide-like repeats (PDB id:1NA0 \citep{regan2003structure}) and right leucine-rich repeats (PDB id:4IM6). The defined repeated unit is highlighted in orange.
}
\label{fig:structures}
\end{figure*}

Repeat proteins are composed of tandem repetitions of similar structural motifs of about 20 to 40 amino acids. Under appropriate conditions, these polymers fold into elongated, non-globular structures (Fig.~\ref{fig:structures}). It is apparent that the overall architecture is stabilized mainly by short range interactions, in contrast to most globular protein domains that usually adopt very intricate topologies \citep{regan2005current}. In their natural context, repeat proteins are frequently found mediating protein-protein interactions, with a specificity rivaling that of antibodies \citep{smerdon1999trends,kajava2001currentstructural,lassle1999bioessays}. Given their structural simplicity and potential technological applications, repeat-proteins are a prime target for protein design, with very successful examples for a variety of topologies  \citep{itzaki2015biochemtrans,baker2015nature,urvoas2010jmbn}. Most of the current design strategies target the creation of rigid native structures with desired folds that, although beautiful, often lose biological functionality \citep{espada2016pressure}. It is becoming clear that the population of `excited states' is crucial for protein function \citep{frauenfelder1987}, and thus tackling energetic inhomogeneities in protein structures may be crucial for understanding how biological activities emerge \citep{ferreiro2014qrb}. The challenge thus relies in finding an appropriate description for the `energy' of each system, a daunting task for large molecular objects such as natural proteins. 

In principle, the natural variations observed for proteins of the same family must contain information about the sequence-structure mapping. A simple model that just takes into account the frequency of each amino acid in each position is insufficient to capture collective effects, yet, for some architectures it is surprisingly good for the synthesis of non-natural repeat-proteins by `consensus' design \citep{regan2003structure,regan2007blabla,pluckthun2003jmb,pluckthun2003jmbLRR}. It is apparent that in the case of repeat proteins the local signals play inordinately large roles in the energy distribution, just as expected from their topology \citep{ferreiro2008energy} and hence, small heterogeneities can be propagated from the local repeat units to higher orders affecting the overall structure and dynamics \citep{parra2013tiling,espada2015biochemtran}. Thus, collective effects may be approximated as small perturbations to local potentials, simplifying the energetic description of complex natural systems \citep{frauenfelder2003myoglobin}.

\medskip
\medskip

In the last years new methods to analyze correlated mutations across a family of proteins have arisen (mfDCA\citep{morcos2011pnas}, plmDCA \citep{aurell2013pre,aurell2014jcp}, Gremlin \citep{balakrishnan2011protein}, EVFold \citep{marks2014elife} to name a few). The main hypothesis behind these methods is that biochemical changes produced by a point mutation should be compensated by other mutations (along evolutionary timescales) to maintain protein viability or function. 
These methods can also be used to disentangle relevant direct correlations from indirect ones. 
They are very successful at predicting spatial contacts and interactions for many protein topologies \citep{morcos2013pnas,cheng2014pnas,onuchic2012pnas,baker2014elife,cheng2016proteinscience}. Nevertheless, these methods do not take into account the chemical nature of the amino acids, which can be codifying inhomogeneities in the energetic distribution that are crucial for the activity of repeat-proteins \citep{pluckthun2007jmb,ladbury2005jmb}. On this basis, different approaches have been proposed recently to include chemical details in the correlation analyses \citep{levy2017potts}, trying to predict folding stability \citep{tiana2015jcp}, conformational heterogeneity \citep{morcos2013pnas,sutto2015pnas,haldane2016protsci} or the global effect on antibiotic resistance from sequences of $\beta$-lactamases \citep{weigt2015mbe,marks2015arxiv}. As many other tools, these were optimized to perform well on globular proteins, and their application to repeat proteins is not straightforward. Besides the point-mutation mechanism, repeat proteins are believed to evolve via duplication and rearrangement of repeats \citep{schuler2016mbe}, resulting in an inherent symmetry which usually confounds sequence analyses \citep{espada2015biochemtran}. Making use of this symmetry, we have previously proposed a specific version of mfDCA and plmDCA for repeat proteins \citep{espada2015BMC}. In this work we develop an alternative `evolutionary field' to approximate the biochemical properties of repeat proteins just from the analysis of natural sequences. We take advantage of the elongated and repetitive structure of these proteins (Fig.~\ref{fig:structures}) to extract as much information as possible from the data, and apply the general ideas on three specific families, ankyrin repeats (ANK), leucine-rich repeats (LRR) and tetratricopeptide-like repeats (TPR). 

\section*{Evolutionary energy for repeat proteins }
To study the co-occurrence of mutations in a sequence alignment of a particular protein family, \cite{weigt2009pnas} proposed a Hamiltonian or energy expression which resembles a Potts model:
\begin{equation}
E(\vec{s}) = - \left[ \sum_{i=1}^L h_i(a_i) + \sum_{i=1}^L\sum_{j=i}^L J_{ij}(a_i,b_j) \right]
\label{eq:E}
\end{equation}
where the set of $\{h_i(a_i)\}$ parameters, one for each amino acid in each position, accounts for a local propensity of having a specific residue on a particular site of the protein, and the set of $\{J_{ij}(a_i,b_j)\}$ indicates the strength of the `evolutionary' interaction between each possible amino acid in every pair of positions along the protein. There are $q=21$ possible values of $a_i$ and $b_j$, one for each amino acid and one for the gaps included on the multiple sequence alignments. This expression is evaluated on a particular sequence on an alignment, and the summations go over the $L$ columns of the alignment. A sequence is more favorable or more energetic if it gets lower values of $E(\vec{s})$. It can be expected that the population of sequences follows a Boltzmann distribution $P(\vec{s})=\frac{1}{Z}e^{-E(\vec{s})}$ \citep{finkelstein1995protein}. The parameters are thus fitted to reproduce the frequencies of occurrence of each amino acid in each position ($f_i(a_i)$) and the joint frequencies of amino acids ($f_{ij}(a_i,b_j)$) in an alignment of natural sequences used as input: 
\begin{eqnarray}
f_i(a_i)=\sum_{a_k, k\neq i} P(\vec{s})\\
f_{ij}(a_i,b_j)=\sum_{a_k, k\neq i,j} P(\vec{s})
\end{eqnarray}
Nevertheless, for repeat proteins there is another feature we want to capture with an evolutionary energy: the high identity of amino acids constituting consecutive repeats, arisen by the repetitiveness of these families and probably a signature of their evolutionary mechanisms (Fig.~\ref{fig:freq}). 
  
Therefore, we propose the following model for repeat proteins: 
\begin{equation}
E(\vec{s}) = - \left[ \sum_{i=1}^L h_i(a_i) + \sum_{i=1}^L\sum_{j=i}^L J_{ij}(a_i,b_j) -\lambda_{Id}(\vec{s})\right]
\label{eq:Eid}
\end{equation}
This expression is designed to be applied in sequences constituted by two repeats. $\lambda_{Id}$ is a parameter that aims at reproducing the probabilities of the percentage of identity (\%Id) between consecutive repeats in natural proteins ($p_{id}$). Basically, it accounts for higher order correlations not captured by the pairwise terms. For a given sequence we calculate the \%Id of the adjacent repeats and sum the parameter $\lambda_{Id}$ corresponding to that \%Id value. When the correct parameters are obtained, this equation can be used to produce an ensemble of sequences consistent with the constraints ($f_i(a_i)$, $f_{ij}(a_i,b_j)$ and $p_{id}$). We work with pairs of repeats as it is the minimum unit that includes the interaction between repeats and the possibility of measure sequence identity between consecutive repeats.
In the following section we will show the convergence of the method and the relevant information that can be obtained from it. For further details about the procedure to assign values to the parameters, please refer to Methods section.

\section*{Results}
\subsection*{\textit{Evolutionary Energy} reproduces ensembles of sequences with natural frequencies and repeat protein characteristics.}

We construct an alignment of pairs of repeats for each family: ANK (PFAM id PF00023, and final alignment of 20513 sequences of L=66 residues each), TPR (PFAM id PF00515, and final alignment of 10020 sequences of L=68 residues each) and LRR (PFAM id PF13516, and final alignment of 18839 sequences of L=48 residues each). See Methods for further details of construction.
We measure $f_{i}(a_i)$, $f_{ij}(a_i,b_j)$ and $p_{id}$. Using a gradient descent procedure we obtain a set of parameters in equation \ref{eq:Eid} which are able to reproduce $f_{i}(a_i)$, $f_{ij}(a_i,b_j)$ and $p_{id}$. 
In principle, the number of parameters is large: $Lq$ $h_i$ parameters, $\frac{(Lq)^2 - Lq}{2}$  $J_{ij}$ parameters and $\frac{L}{2}+1$ $\lambda_{Id}$. For example, for pairs of ANK repeats this means 1386 $h_i$, 959805 $J_{ij}$ and 34 $\lambda_{Id}$. To reduce the number of free parameters to fit we use a $L_1$-regularization which fixes to zero those parameters which do not contribute significantly to fit the frequencies. This regularization allows us to set to exactly zero between 85 and 91\% of the $J_{ij}$ parameters which, when they are free to vary, only reach small values (Suppl. Fig.~\ref{fig:jijReg}). We bound the maximum error permitted in the frequency estimations to 0.02. Refer to Methods for more details.

In the three families studied, the parameters obtained allow us to generate ensembles of sequences which reproduce natural $f_{i}(a_i)$, $f_{ij}(a_i,b_j)$ and $p_{id}$ (Fig.~ \ref{fig:freq}A). Notice that most frequencies are fitted with an error of an order of magnitude lower than the maximum bound imposed (Suppl. Fig.~\ref{fig:histerrors}).

\medskip
\begin{figurehere}
\centering
\includegraphics[width=.5\textwidth]{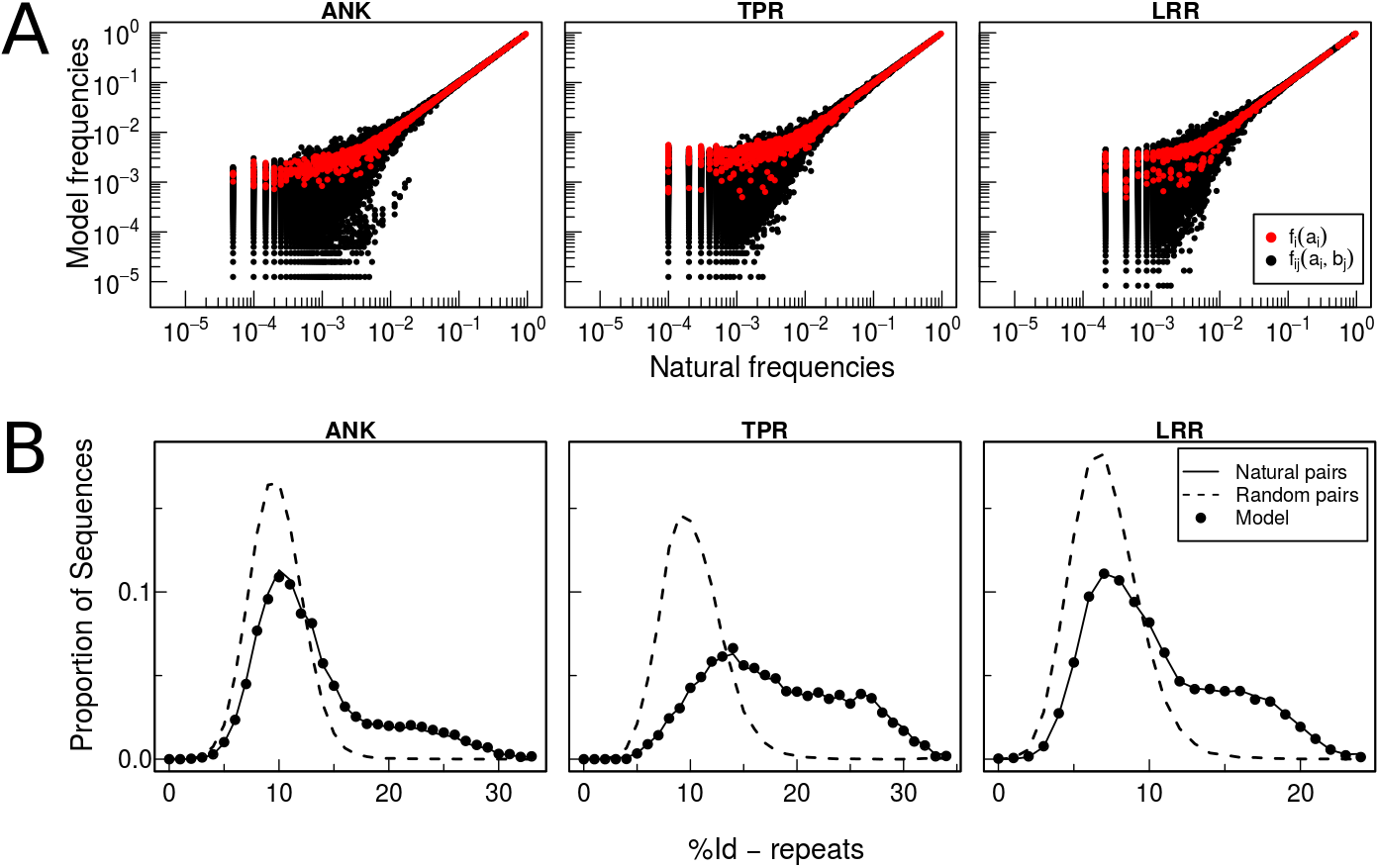}
\caption{The proposed model fits the frequencies of amino acids and natural repeat-identities $p_{id}$. On A, we compare  marginal frequencies $f_i(a_i)$ (red) and joint frequencies $f_{ij}(a_i,b_j)$ (black) on the natural ensemble of sequences (x-axis) and on the set of sequences generated by the model (y-axis). On B, we calculate the distribution of identity between repeats $p_{id}$ for consecutive repeats (solid line), and for natural repeats which are not consecutive, i.e. they are not next to each other in the primary structure (dot lines). Consecutive repeats present a population with high identity between repeats that any pairs of repeats do not show. We compare the distribution produced by the model $p_{id}^{model}$ (dots).}
\label{fig:freq}
\end{figurehere}
\medskip

The $p_{id}$ distributions are also very well reproduced (Fig.~\ref{fig:freq}B). Not only the general shape, but also the populated long tail for highly similar repeats. It is not possible to obtain the same distribution only by fitting amino acid frequencies $f_i(a_i)$ and $f_{ij}(a_i,b_j)$, it is mandatory to explicitly include the $p_{id}$ by including the parameters $\lambda_{Id}$ (Suppl. Fig.~\ref{fig:pidLambda}), suggesting that higher order correlations must be accounted for describing these systems.

\subsection*{Evolutionary Energy distinguishes between proteins on a given family and other polypeptides}
Once the set of parameters \{$h_i(a_i)$,$J_{ij}(a_i,b_j)$,$\lambda_{Id}$\} is obtained, it can be used to score any sequence of $L$ amino acids via equation \ref{eq:Eid}. In this section we test if this measure is capable of distinguishing polypeptides that fold in a three dimensional structure similar to members of the repeat protein family from those that do not. 

\begin{figure*}
\centering
\includegraphics[width=.9\textwidth]{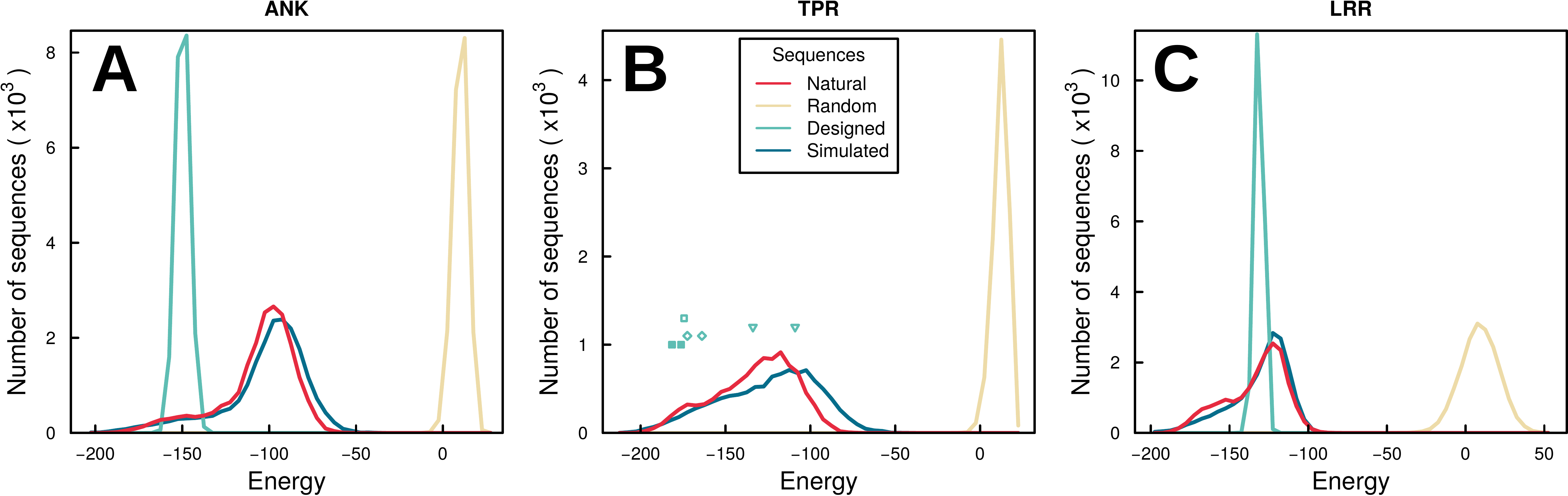}
\caption{Energy score distribution for different ensembles of sequences. Red lines, natural sequences used to train the model on equation \ref{eq:Eid}. Blue lines, sequences simulated by Monte Carlo under expression \ref{eq:Eid}. In the three families, it overlaps with natural sequences, suggesting that simulated sequences imitate the natural ensemble. Yellow lines, strings of random amino acids used as negative control. They show that the energy distinguishes between polypeptides belonging to a protein family and other strings of amino acids. Green lines, squares, diamonds and triangles, energies for designed proteins.}
\label{fig:Hist}
\end{figure*}

We calculate the distribution of energies of different sets of sequences (Fig.~\ref{fig:Hist}). The ensembles of natural sequences of each protein family used to learn the parameters have a unimodal distribution of energies centered around -100 (Fig.~\ref{fig:Hist}, dots).
These distributions are clearly differentiated from the energies of random chains of residues (Fig.~\ref{fig:Hist}, red lines). 

\medskip
\medskip

For a positive control we evaluate designed proteins which have been experimentally synthesized. For the ANK family, we consider the library of repeat sequences built by Pl\"uckthun's laboratory \citep{pluckthun2003jmb} (blue dots line, Fig.~\ref{fig:Hist}A). This library was constructed by fixing on each repeat 26 positions out of 33 to the most frequent residue in the multiple sequence alignment. This resulted in a set of sequences that have small variations with respect to the ANK consensus (the sequence with the most frequent amino acid in each position). In our expression, they score a very low energy distribution, overlapping with the most negative tail of the distribution of natural sequences. It is notable that consensus designed ANK have been shown experimentally to be extremely stable.
For the TPR family, consensus designed was done by Regan's laboratory \citep{regan2007blabla,regan2003structure}. All pairs of repeats synthesized have the same amino acid sequence, and it's energy score is indicated by a green full square in Fig.~\ref{fig:Hist}B. Again, the designed sequence matches values at the most left side of the energy distribution of natural sequences, and coincidentally reports high folding stability. From it, other variants with few point mutations to improve binding to a specific ligand have been synthesized. As shown in empty green squares \citep{krachler2010} and diamonds \citep{cortajarena2010} in Fig.~\ref{fig:Hist}B, they have higher energy, but still in the left most side of natural sequences distribution. Recently, a different design strategy was done \citep{lupas2016}. Based on a non-repetitive protein, but similar to TPR fold, they put togheter various repetitions of the fold, using TPR loops to link them. They obtained a three-repeats protein whose pair of repeats energy are represented on triangles on Fig.~\ref{fig:Hist}B. This time, they match natural sequences distribution in higher values.

Finally, for the LRR family we contrast with the library of proteins designed by Pl\"uckthun's group based on the consensus sequence \citep{pluckthun2003jmbLRR}. The repetition they considered has 57 amino acids, which includes two types of repeats, one of 28 residues and the other one of 29. As the repeat we are using for LRR is 24 residues long, we aligned both definitions and evaluated the library removing the amino acids not matching our definition. Again, their scores form a narrow distribution, but this time it is not placed on the most favorable side of the natural sequences distribution (Fig.~\ref{fig:Hist}C). Coincidentally, selected species studied do not show such a high folding stability as the ANK library did.

With these parameters, we are able to generate an ensemble of sequences which are in agreement with the constraints used, via a Monte Carlo simulation (see Methods). The distribution of energies of these simulated sequences matches the natural sequences energies distribution with remarkable accuracy. Moreover, we randomly choose 100 sequences from the natural ensemble and 100 sequences from the simulated one, perform a Smith-Waterman pairwise alignment all against all, calculate the pair similarity using BLOSUM62 matrix and used it as a distance method to plot a dendogram of the sequences (Supl. Fig.~\ref{fig:dendograms}). Both species appear interspersed, showing that it is not possible to distinguish a natural sequence from a constructed one. Also, we tested  \textit{familiarity} to the ANK family as defined in \cite{turjanski2016protein}  and found overlapping distributions for both species (Supl. Fig.~\ref{fig:familiarity}).
Therefore, simulated sequences represent possible variants to natural repeats. The wide distribution of natural proteins suggests that it should be possible to engineer sequences with more variable repeats, more dissimilar among neighbors and to the consensus than the ones published up to date.

\subsection*{Low evolutionary energy sequences have similar repeats}
Are there any invariant properties shared by low energy sequences? Given that repeat-proteins may evolve by other mechanisms besides point substitutions, we analyze if low energy sequences are constituted by highly similar repeats and if they are close to consensus sequences.

On Fig.~\ref{fig:pid}A we show the relation between the \%Id between the repeats and the energy of the sequence. It is evident that low energy sequences are constructed by pairs of highly similar repeats. This could be a transitive effect: if low energy sequences are very similar to the consensus sequence, and the consensus sequence is formed by two identical repeats, we would be seeing that more similarity between repeats causes lower energies. We can see that it is not the case (Fig.~\ref{fig:pid}B). We plot the \%Id to the consensus against the energy of each sequence. The consensus was calculated with the most frequent amino acid in each position on sequences used as input. We can see that there is no evident correlation between the energy and the similarity to the consensus. Thus, low energy sequences that differ from the consensus one may be constructed. Also, there are no sequences which get a high \%Id to the consensus. We conclude that there are different repeats which have low energies within a protein family, and not only the consensus sequence.

\medskip
\begin{figurehere}
\centering
\includegraphics[width=.5\textwidth]{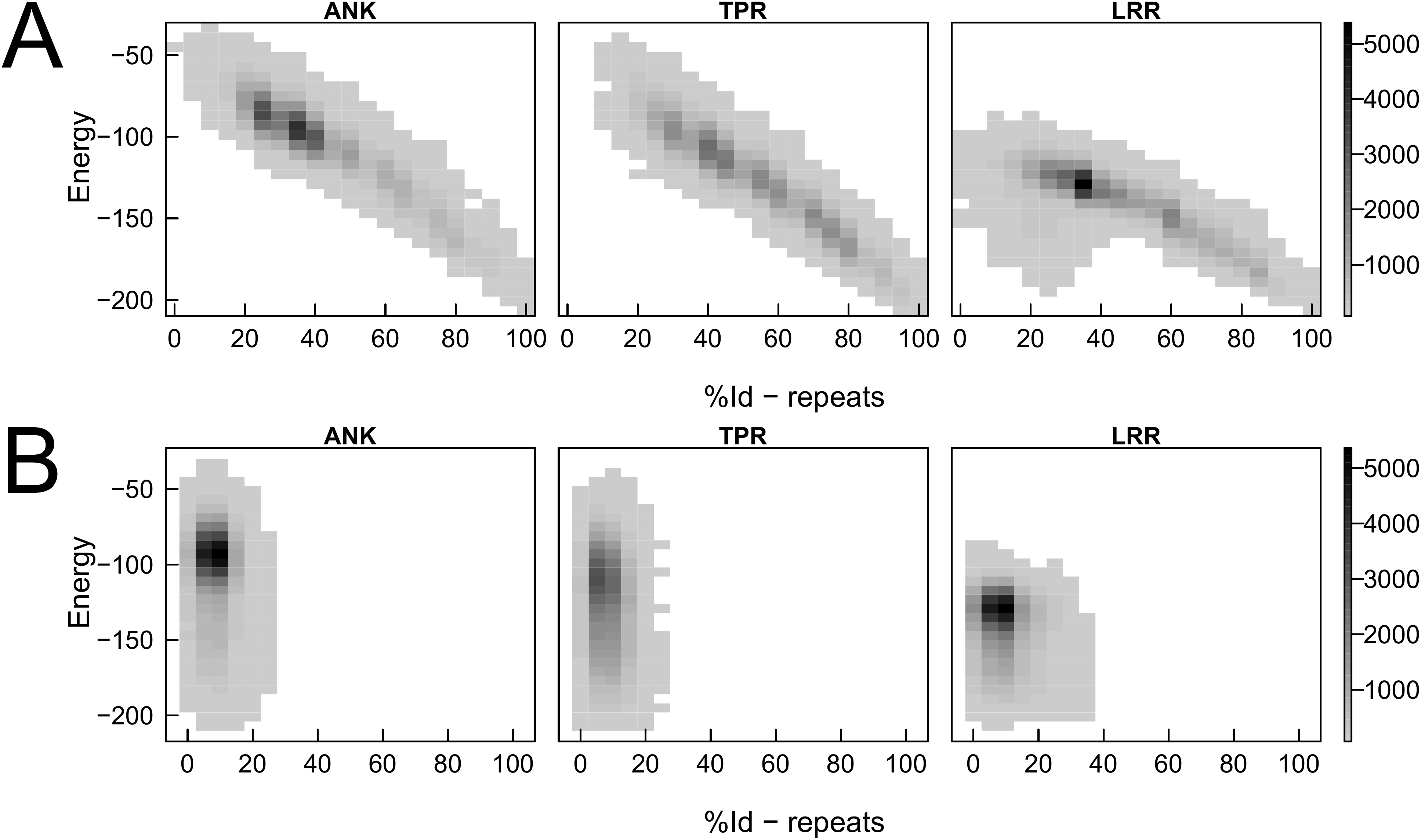}
\caption{Most favorable simulated sequences have very similar repeats, yet they are different to the consensus repeat. On A, we plot the energy vs. the identity between the repeats that constitute the sequence. Even though the deviation is large, most stable sequences tend to have more similar repeats. On B, we plot the energies of simulated sequences vs the identity to the consensus of the family. In all cases, the identity to the consensus is small and uncorrelated to the energy, indicating that sequences which differ significantly from the consensus can be stable variants of the family. }
\label{fig:pid}
\end{figurehere}
\medskip

\subsection*{Evolutionary Energy and folding stability change upon point mutations}

Consensus designed ANK proteins are very stable upon chemical and thermal denaturation \citep{pluckthun2003jmb}, and, as shown in Fig.~\ref{fig:Hist} also score a very low evolutionary energy according to equation \ref{eq:Eid}. Can we quantify the relationship between the stability and the evolutionary energy? 

A potential test can be performed by comparing to experiments in which the effect of point mutations was evaluated. These incorporate one, two or three point mutations in natural proteins, and characterize the unfolding free energy $\Delta$G of the wildtype and the mutated variant. A higher $\Delta$G reports a more stable protein. We compare the change in the $\Delta$G between the mutated and the wildtype protein ($\Delta\Delta$G), and the difference of energy for their sequences according to equation \ref{eq:Eid}.

Although the energy expression is learned for pairs of repeats, we can easily extend it to an array of repeats making use of the elongated structure of repeat proteins in which only adjacent repeats interact. From our expression we have parameters assigned to intra-repeat positions ($h_i$ with $i = 1\dots \frac{L}{2}$ and $J_{ij}$ with $i,j$ $=1\dots \frac{L}{2}$), and inter-repeat interactions ($J_{ij}$ with $i = 1\dots \frac{L}{2}$ and $j=\frac{L}{2}+1 \dots L$, and $\lambda_{Id}$). Then for each repeat we can assign an internal energy $\sum_{i=1}^{L/2} h_i(a_i) + \sum_{i=1}^{L/2}\sum_{j>i}^{L/2} J_{ij}(a_i,b_j)$ and a interaction energy $\sum_{i=1}^{L/2}\sum_{j= L/2+1}^L  J_{ij}(a_i,b_j)+\lambda_{Id}$, which of course depends on the amino acids constituting each repeat. 

On Fig.~\ref{fig:DDG}A, we show the comparison between $\Delta\Delta$G and the evolutionary energy calculated using Eq.~\ref{eq:Eid}, 
done for three different ANK proteins: I$\kappa$B$\alpha$ \citep{devries2011jmb,ferreiro2007jmb}, Notch \citep{street2005improved} and p16 \citep{tang2003sequential}. It should be noted that different experimental techniques return different values for $\Delta$G for the same protein, non overlapping within experimental error, pointing that other factors contribute to the experimental quantification of $\Delta\Delta$G. A linear fit returns $R^2 \approx 0.49$. Nevertheless, from 152 mutations we analyzed, 114  (75 \%) are predicted favorable when the mutation stabilized the folding of the structure, and unfavorable when they have also been measured to destabilize. The predictions that deviated the most are mutations in Notch from Serine to Proline, which is a structural disruptor, and were not considered in the linear fit.

\medskip
\begin{figurehere}
\centering
\includegraphics[width=0.5\textwidth]{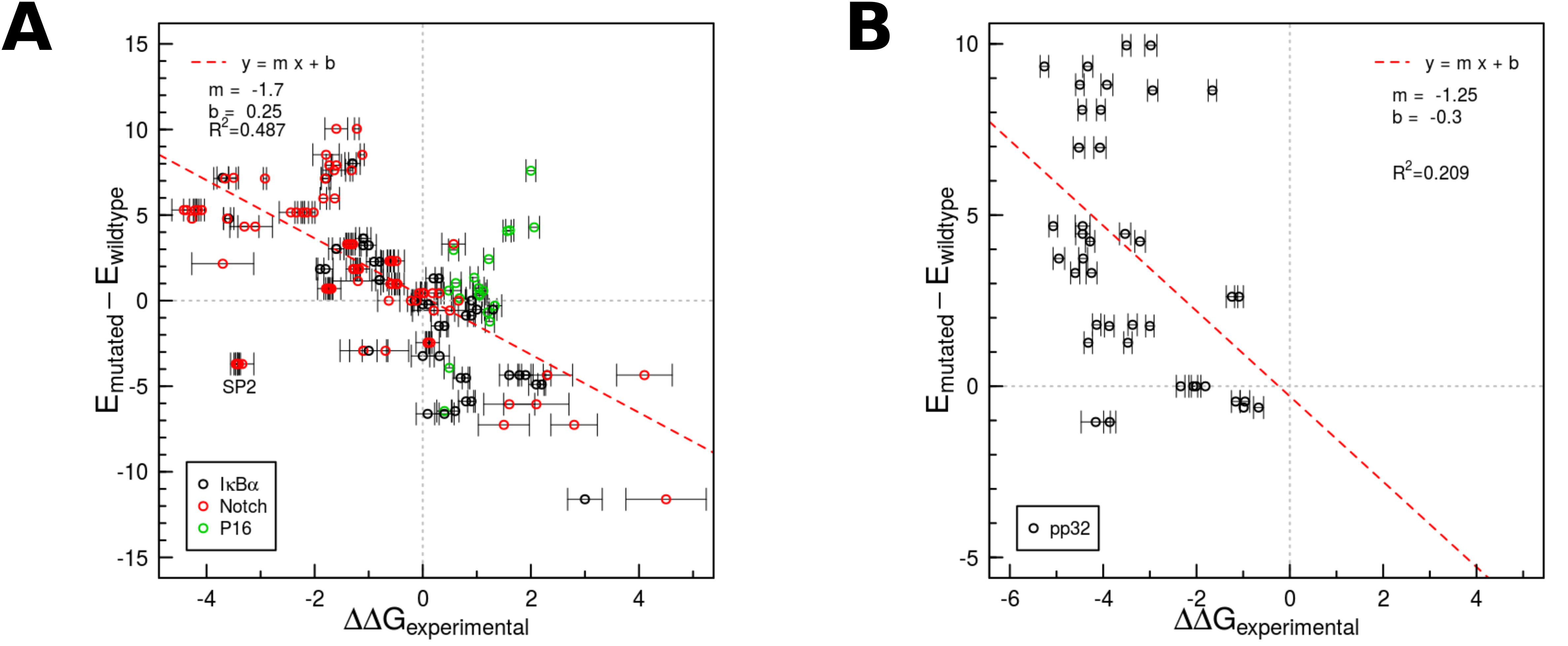}
\caption{Variation of energy score as a predictor of the folding stability upon point mutations. We compare difference in unfolding $\Delta$G between a wildtype protein and a mutated variant (x-axis) and the change in energy according to Eq. \ref{eq:Eid}. Error bars indicate the experimental standard deviation. On A, for proteins belonging to ANK family, and on B for LRR.}
\label{fig:DDG}
\end{figurehere}
\medskip

On Fig.~\ref{fig:DDG}B, we show reported mutations on pp32 \citep{barrick2015pnas}, a protein belonging to LRR family. Again, measurements with different methods report different values of $\Delta\Delta$G. The linear fit returns a poor $R^2 \approx 0.21$, but 30 (75\%) mutations are both predicted and reported unstabilizing. 
  
A similar comparison was performed by \cite{tiana2015jcp} for small globular proteins with an expression related to Eq. \ref{eq:E}. To reduce the number of interaction parameters $J_{ij}(a_i,b_j)$ they explicitly used structural information and set to zero all interactions between positions which are not in contact in the native structure. In contrast, we use a $L_1$-regularization to fix to zero those parameters which do not contribute significantly to the fitting process and obtain $J_{ij}(a_i,b_j)=0$ and $J_{ij}(a_i,b_j)\neq 0$ in all pairs of positions, regardless they are supposed to be in contact or not in the 3D structure.

\subsection*{Interaction parameters are related to the structure and the sequence symmetry}

Are the obtained parameters related to structural properties of these proteins? 
Local fields, $h_i(a_i)$, should account for the local propensity of each amino acid in each position, and therefore are expected to be related to $f_i(a_i)$. 
Fig.~\ref{fig:campo1}A shows that the inferred $h_i(a_i)$ parameters are different from the initial condition $-\ln(f_i(a_i))$ 
for the ANK family; that is, the values obtained for the parameters that account for higher order correlations are relevant. In red we highlight the points related to the consensus amino acid in each position. All of these residues have a strong local field associated to them, justifying why the construction of sequences with these amino acids results in foldable proteins.
We also show a contact map of two ANK repeats (PDB id: 1N0R) on Fig.~\ref{fig:campo1}B: gray background indicates that the two positions given by x and y axis are in contact in the native structure, and white that they are not. On the upper triangle of the figure and in blue crosses, we mark the positions involved in the highest $J_{ij}$ parameters, i.e. those which imply higher coupling. A darker blue indicates that there are more $J_{ij}$ (more combinations of amino acids) between those positions. Most of the highest $J_{ij}$ match a pair of positions in contact in the 3D structure, or two which correspond to the same residue in the adjacent repeat patterns, i.e.  i-th position in the first repeat and position j=i+33 in the second repeat. In red crosses we show the lowest $J_{ij}$, that mark a negative constraint. Again, a darker red means that there are more $J_{ij}$ with low values between those positions. It is apparent that these also involve mostly residues in contact, but shows that other regions are responsible for negative design.

\medskip
\begin{figurehere}
\centering
\includegraphics[width=.5\textwidth]{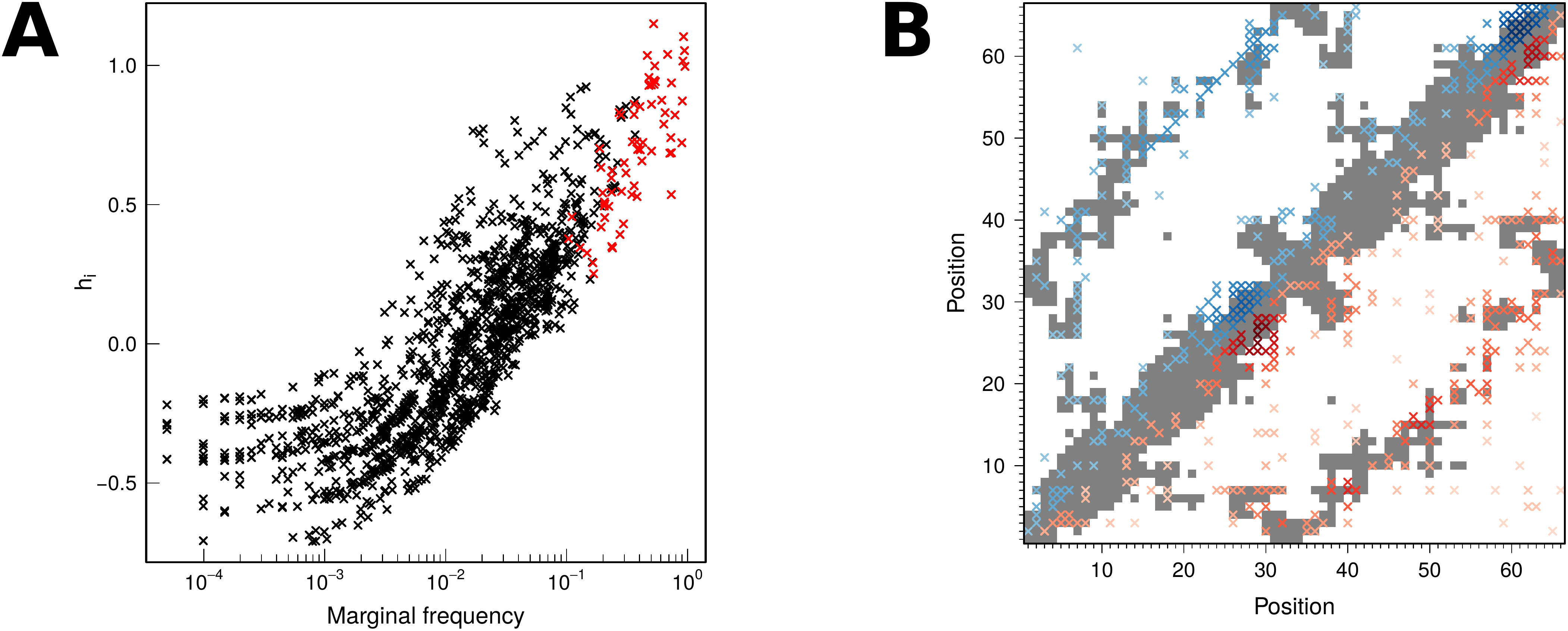}
\caption{For the ANK family, on panel A we compare the parameters $h_i(a_i)$ to the marginal frequencies. The site-independent model (and initial condition) states that $h_i(a_i) = \ln(f_i(a_i))$. For the final model, this relation is tuned by the higher order correlations. On red, the parameters associated with the most common amino acid in each position are highlighted. On panel B, we compare the contact map of a pair of repeats of 1N0R (gray shadow) and the highest (blue) and lowest (red) $J_{ij}(a_i,b_j)$ parameters. The color scale indicates how many parameters involves the two positions (due to different sets of amino acids). Most extreme values fall into residues in contact or in the equivalent position of a repeat.}
\label{fig:campo1}
\end{figurehere}
\medskip

\section*{Discussion}

We propose a statistical model to account for fine details of the energy distribution in families of repeat proteins using only the sequences of amino acids. The model consists of a generalization of a Potts model to account for the local and pair-wise interactions and an extra term that includes higher order correlations, accounting for the similarity between consecutive repeats. The model is constrained by evolutionary characteristics of the families of proteins: we measure the frequencies of amino acids, co-occurrence of amino acids and the identity between repeats in extant natural proteins. To statistically define these quantities it is necessary to have a large set of sequences, which we showed are currently available for several repeat-protein families \citep{espada2015BMC}. No information about the native folded conformation is required. The computation of the evolutionary energy field is computationally demanding, mostly due to long times spent in rigorous Monte Carlo simulations, but once the fitting is done the parameters can be used to score individual sequences fast and easily. 

We studied three popular repeat protein families: ANK, TPR and LRR. After pre-processing of the alignments, we had enough sequences ($\approx$ 20500, 10000 and 18800 respectively) to fit the model to pairs of repeats of each family. We scored the \textit{evolutionary energy} of all natural sequences in PFAM, and it allowed us to clearly distinguish between natural proteins and random sequences of amino acids: the first have energy values $<$ -50 and show a large spread while all random sequences have energy values $\approx$ 0. We evaluated designed repeat proteins which have been shown to fold and found that they score within the natural sequences distribution of energies. For the ANK and TPR family, these designed proteins have been shown to be highly stable upon thermal and chemical denaturation and, coincidentally, they are located at the most favorable side of the energy distribution of natural proteins, suggesting that the evolutionary energy score can be related to folding stability.

The energetic model can be used in Monte Carlo simulations to generate sequences that agree with the natural constraints of a given protein family. This ensemble of simulated sequences matches the amino acid frequencies, the identity between repeats and also the energy distribution of natural proteins. We found this set of simulated sequences is statistically indistinguishable from natural counterparts. Thus, the proposed model can be used as a tool to design repeat-protein sequences that have all the natural characteristics evaluated to date. Moreover, the stability change upon single point mutation can be well predicted by the model. For both the simulated sequences and for natural counterparts, we found that the similarity between consecutive repeats correlates with lower energy values, and that these are not necessarily similar to the consensus sequence of the family, pointing out that duplication of stretches of sequences may well be an important factor in the evolution of these systems \citep{bjorklund2006ploscompbiol}. 

The existence of a simple and reliable energy function to score the `evolutionary energy' of repeat-proteins can be used to trace the biological forces that acted upon their history, and to explore to which extent these conflict with the physical necessities of the systems \citep{morcos2014tsel}. Mapping the energy inhomogeneities along the repeat-arrays may allow us to infer the population of excited states in these proteins, many of which have been related to their physiological mechanisms.

\section*{Methods}

\subsection*{Sequence alignments}
Sequences of repeats were obtained from PFAM 27.0 \citep{bateman2004pfam}. These sequences usually have misdetected initial and final residues. We completed these positions with the amino acids present on the actual proteins. This leads to a reduction on the number of gaps in our alignments, which usually derives into noisy predictions in correlation analyses \citep{tiana2015jcp}. After, we created the alignment of pairs of repeats, joining sequences of repeats which are consecutive in a natural protein. Finally, we removed insertions from the alignments by deleting positions which have gaps in more than 80\% of the sequences in the alignment.

\subsection*{Frequency calculations}
Our model fits the occurrence of amino acids in every position, which we call the marginal frequency of residue $a_i$ at position $i$ of the alignment and denote $f_i(a_i)$, and the joint occurrence of two amino acids $a_i$ and $b_j$ simultaneously at two different positions of the alignment, $f_{ij} (a_i,b_j)$. To avoid biases by the overrepresentation of some proteins in the database, we used CD-HIT \citep{cdhit2002} to choose representative sequences which differ between them in more than 90\% of identity percentage. Finally, we computed by counting the $f_i(a_i)$ and $f_{ij}(a_i,b_j)$, and divided by the total number of sequences.

\subsection*{$p_{id}$ calculations}
From the same alignment explained in \textit{Frequencies calculations}, for a sequence which has $L$ residues constituting two consecutive repeats, the \%Id between the repeats is the number of amino acids in positions $i$ and $i+\frac{L}{2}$, for $i=1\dots \frac{L}{2}$ which are exactly the same. Gaps are treated as an amino acid. Once we have the values for all sequences in an alignment, we define $p_{id}$ as the proportion of sequences within the alignment with the same $\%Id$ between repeats.

\subsection*{Construction of an ensemble of sequences in agreement with a energy equation}
Given a set of parameters ${h_i,J_{ij},\lambda_{Id}}$ and Eq. \ref{eq:Eid}, we use a Monte Carlo procedure and the Metropolis criterion to generate an ensemble of N sequences of length L each. We initiate with a random string of L residues. At each step, we produce a point mutation in any position. If this mutation is favorable, i.e. the energy is lower than that of the original sequence, we accept the mutation. If not, we accept the mutation with a probability of $e^{-\Delta E}$, where $\Delta E$ is the difference of energy between the original and the mutated sequence. When accepted, the mutated sequence is used as the original one for next step. We add one sequence to our final ensemble every t steps (we used t=1000). 

\subsection*{Learning the parameters for the model}
Our model is proposed to reproduce $f_i(a_i)$, $f_{ij}(a_i,b_j)$ and $p_{id}$ from the alignment of natural sequences. To learn the set of parameters ${h_i,J_{ij},\lambda_{Id}}$ which reproduce them, we used a gradient descent procedure. 
In each step, an ensemble of N=80000 sample sequences was produced via Monte Carlo using as energy the expression \ref{eq:Eid} and the trial parameters. We measured its marginal, joint frequencies and $p_{id}$ and updated the local parameters according to:
\begin{equation}
h_i^{t+1} \leftarrow h_i^t - \epsilon_s \left[f_i(a_i) - f_i^{model}(a_i)\right]
\label{eq:h}
\end{equation}
As the number of parameters for coupling is large ($=21^2 L^2$), we used a regularization $L_1$ to force to 0 those parameters which are not contributing significantly to the modeled frequencies. Then, we update  these parameters by:
\footnotesize
\begin{equation}
J_{ij}^{t+1} \leftarrow \left\lbrace 
	\begin{array}{ll}
    	  0 & \text{if } J_{ij}^t=0 \text{ and } \Delta < \gamma \\
      \epsilon_j \left[ - \gamma sign (\Delta)\right]  & \text{if } J_{ij}^t=0 \text{ and } \Delta > \gamma \\
      J_{ij}^t + \epsilon_j \left[\Delta-\gamma sign(J_{ij}^t) \right] & \text{if } \left[ J_{ij}^t + \epsilon_j (\Delta -\gamma sign(J_{ij}^t)\right]\cdot J_{ij}^t >0 \\
      0 & \text{if } \left[ J_{ij}^t + \epsilon_j (\Delta -\gamma sign(J_{ij}^t)\right]\cdot J_{ij}^t <0 \\
	\end{array} \right.
    \label{eq:j}
\end{equation}
\normalsize
where $\Delta=f_i(a_i) - f_i^{model}(a_i)$. Finally, the parameters $\lambda_{Id}$ are updated according to:
\begin{equation}
\lambda_{Id}^{t+1} \leftarrow \lambda_{Id}^{t} + \epsilon_{ID} \left[ p_{id}(\%Id) - p_{id}^{model}(\%Id)\right]
\label{eq:lambda}
\end{equation}

We iterated until the maximum difference between the predicted frequencies and the natural sequences was below 0.02.
The code was written in C++ and is available at GitHub.

\section*{Acknowledgments}

This work was supported by Universidad de Buenos Aires and Consejo Nacional de Investigaciones Científicas y Técnicas CONICET (ANPCyT grant PICT2012-1647 to R.E. and D.U.F.). Work in Paris was supported by grant ERCStG n. 306312.  R.G.P. is a long-term EMBO postdoctoral fellow. R.E. and R.G.P. would like to thank C.F.K.

\bibliographystyle{unsrtnat}
\bibliography{bibliography}

 \end{multicols}

\newpage
\setcounter{figure}{0} 
\setcounter{page}{1} 

\begin{center}
\large
Inferring repeat protein energetics from evolutionary information\\
\vspace*{2pt}
Espada R, Parra RG, Mora T, Walczak AM, Ferreiro DU
\end{center}
 \normalsize
 
\section*{Supplementary information}

\medskip
\begin{figurehere}
\centering
\includegraphics[width=.5\textwidth]{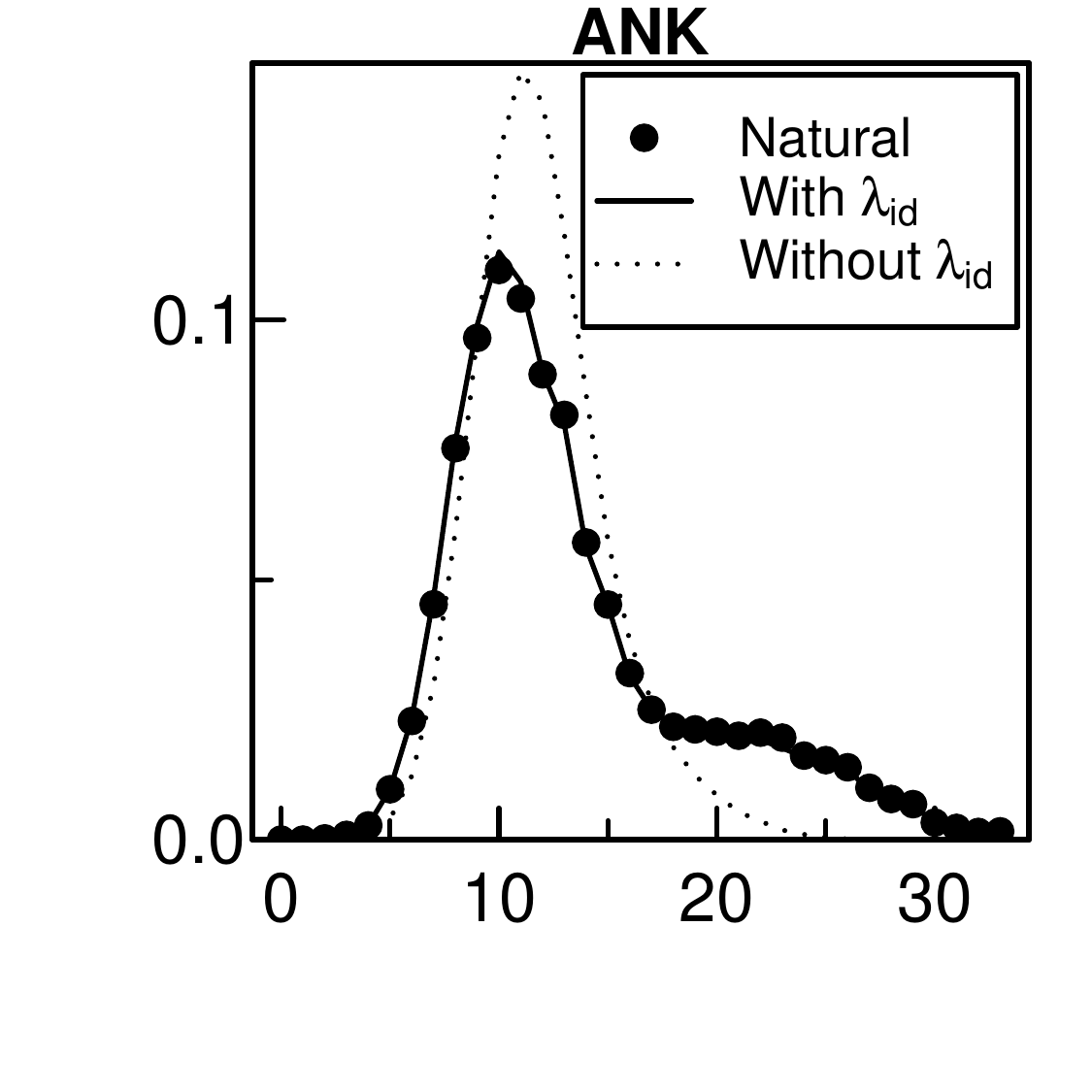}
\caption{Distribution of \% Id of sequences generated by model on equation \ref{eq:E} (main text) in red solid lines and by model \ref{eq:Eid} (main text) in black solid lines. In black dots, the natural sequences' distribution of \%Id.}
\label{fig:pidLambda}
\end{figurehere}
\medskip

\medskip
\begin{figurehere}
\centering
\includegraphics[width=.65\textwidth]{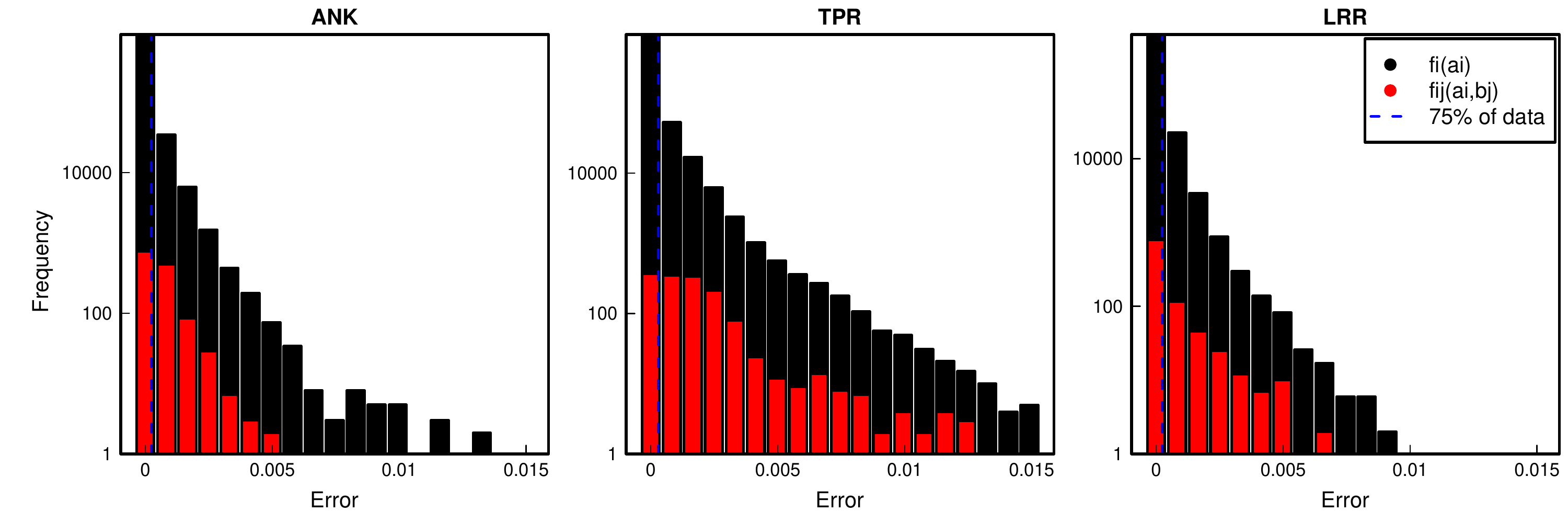}
\caption{Histogram of frequency errors.}
\label{fig:histerrors}
\end{figurehere}
\medskip

\medskip
\begin{figurehere}
\centering
\includegraphics[width=.6\textwidth]{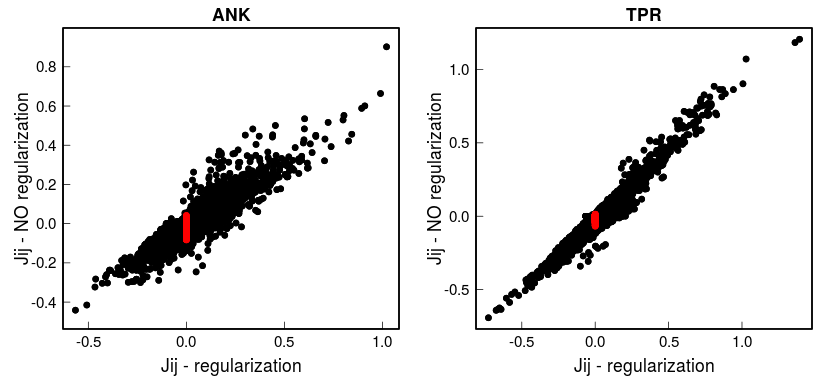}
\caption{Change of $J_{ij}$ parameters under regularization.}
\label{fig:jijReg}
\end{figurehere}
\medskip

\newpage
\medskip
\begin{figurehere}
\centering
\includegraphics[width=.7\textwidth]{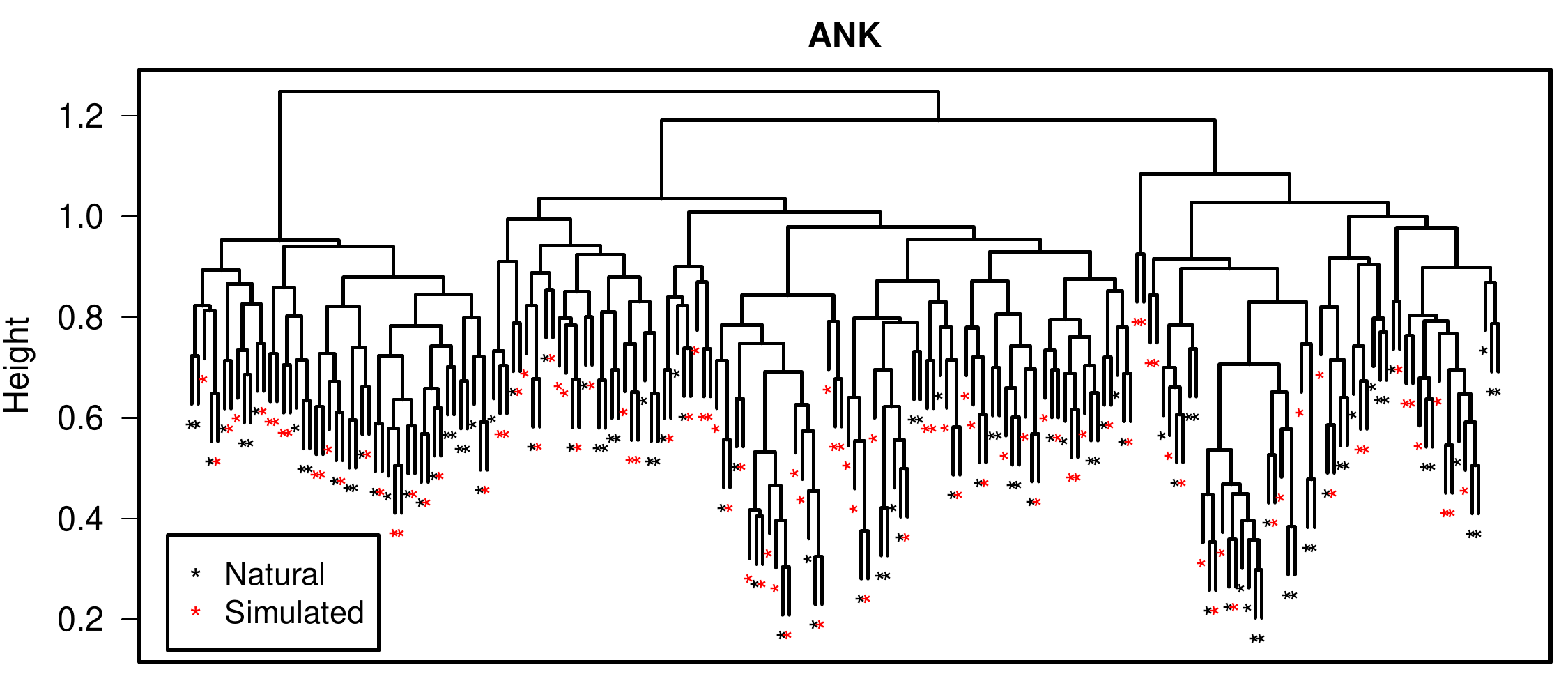}

\includegraphics[width=.7\textwidth]{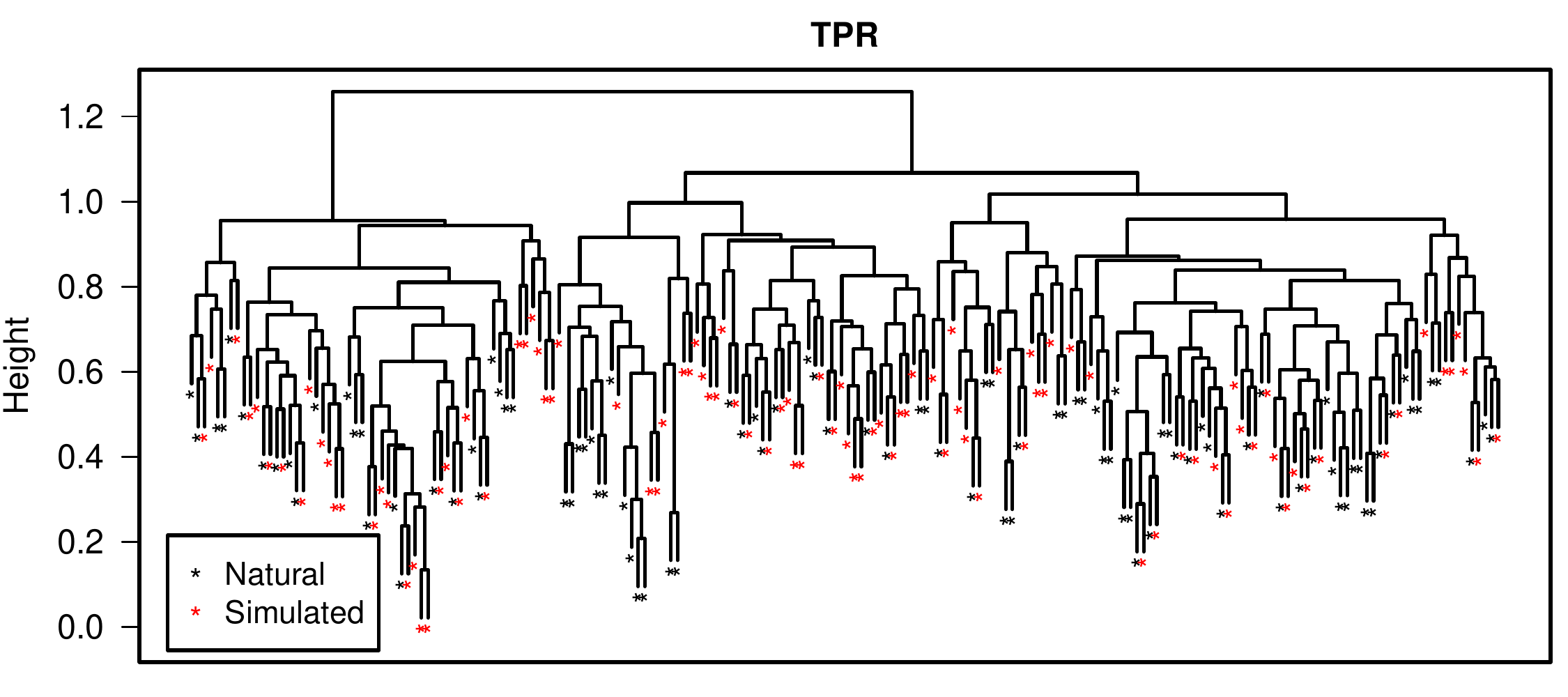}

\includegraphics[width=.7\textwidth]{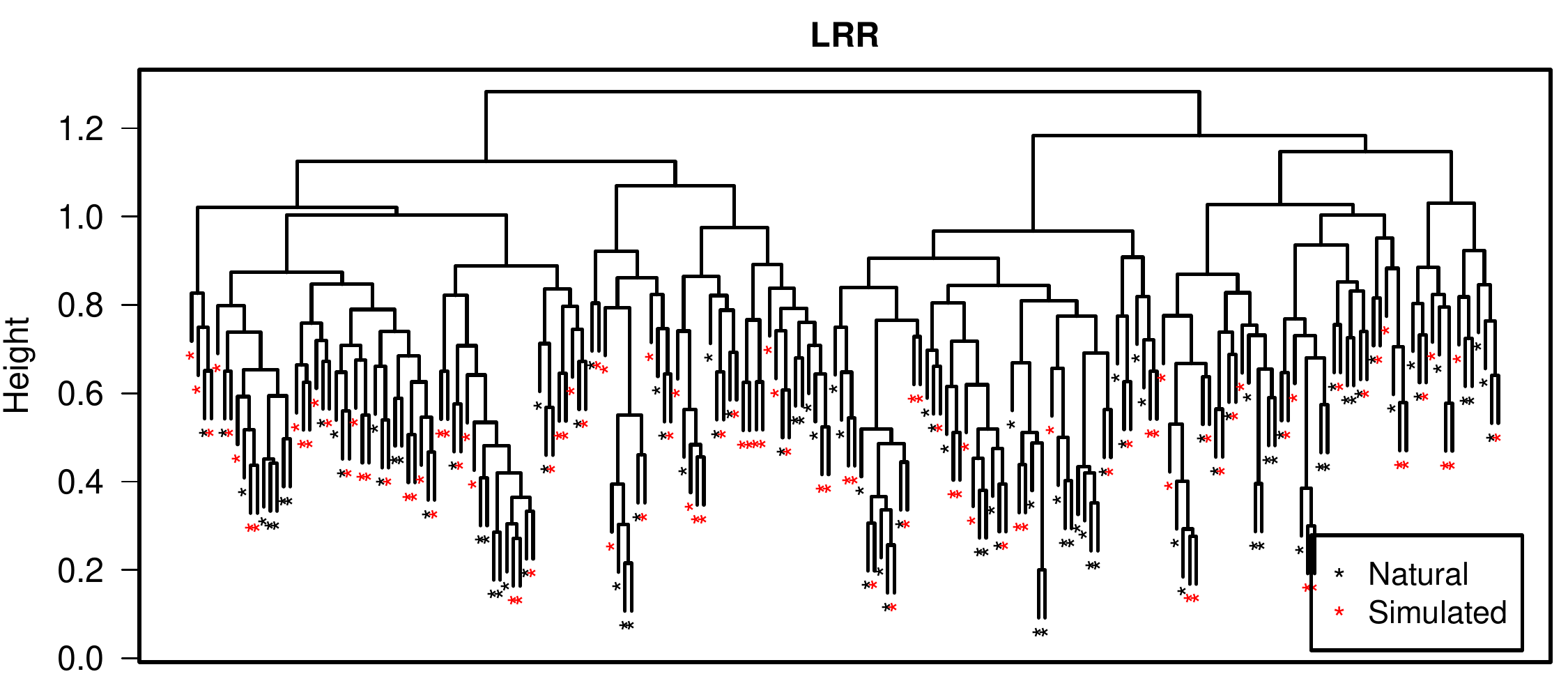}
\caption{Dendogram based on pair-wise similarity. Natural and simulated sequences are indistinguishable from pairwise similarity.}
\label{fig:dendograms}
\end{figurehere}
\medskip

\newpage
\medskip
\begin{figurehere}
\centering
\includegraphics[width=.9\textwidth]{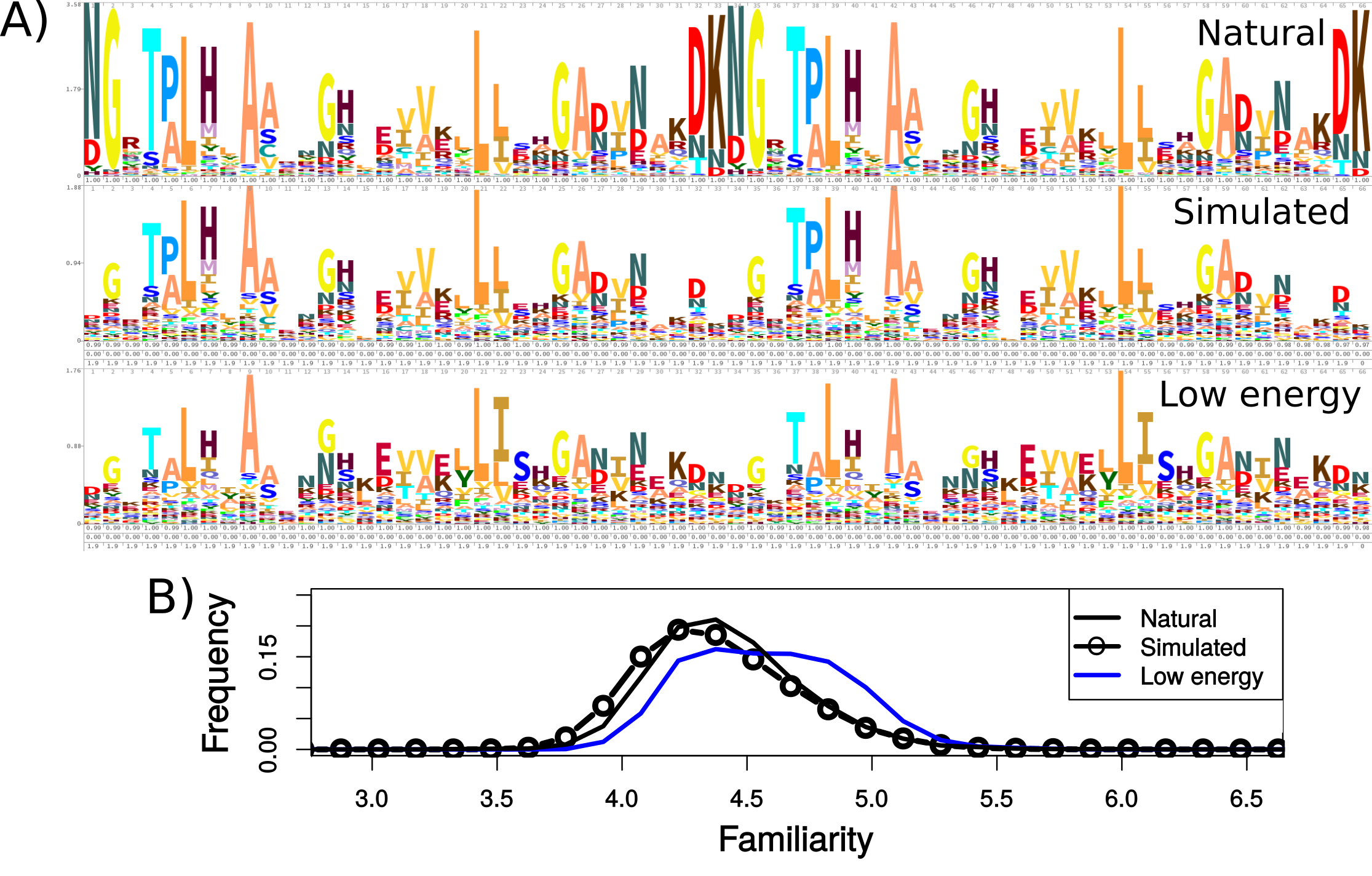}

\caption{A) Logos for the MSA of natural pairs of ANK repeats (top), of simulated pairs of ANK repeats (center) and low energy pairs of simulated ANK repeats (bottom). B) Distribution of familiarity, as defined in Turjanski et al (2016) for the same sets of sequences. Simulated sequences reproduce the distribution of Natural proteins and are indistinguishable. }
\label{fig:familiarity}
\end{figurehere}
\medskip

\newpage

\medskip
\begin{figurehere}
\centering
\includegraphics[width=.3\textwidth]{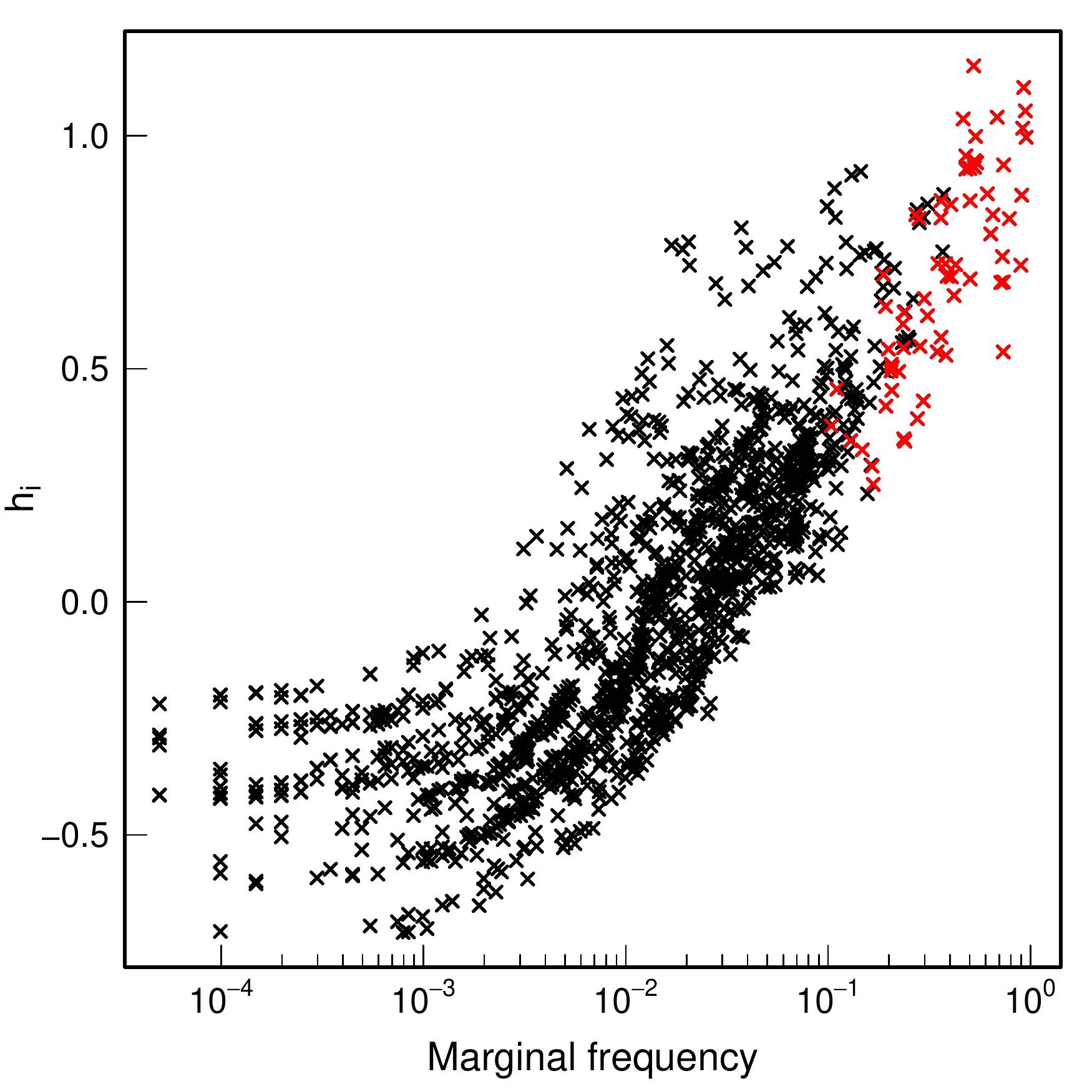}
\includegraphics[width=.3\textwidth]{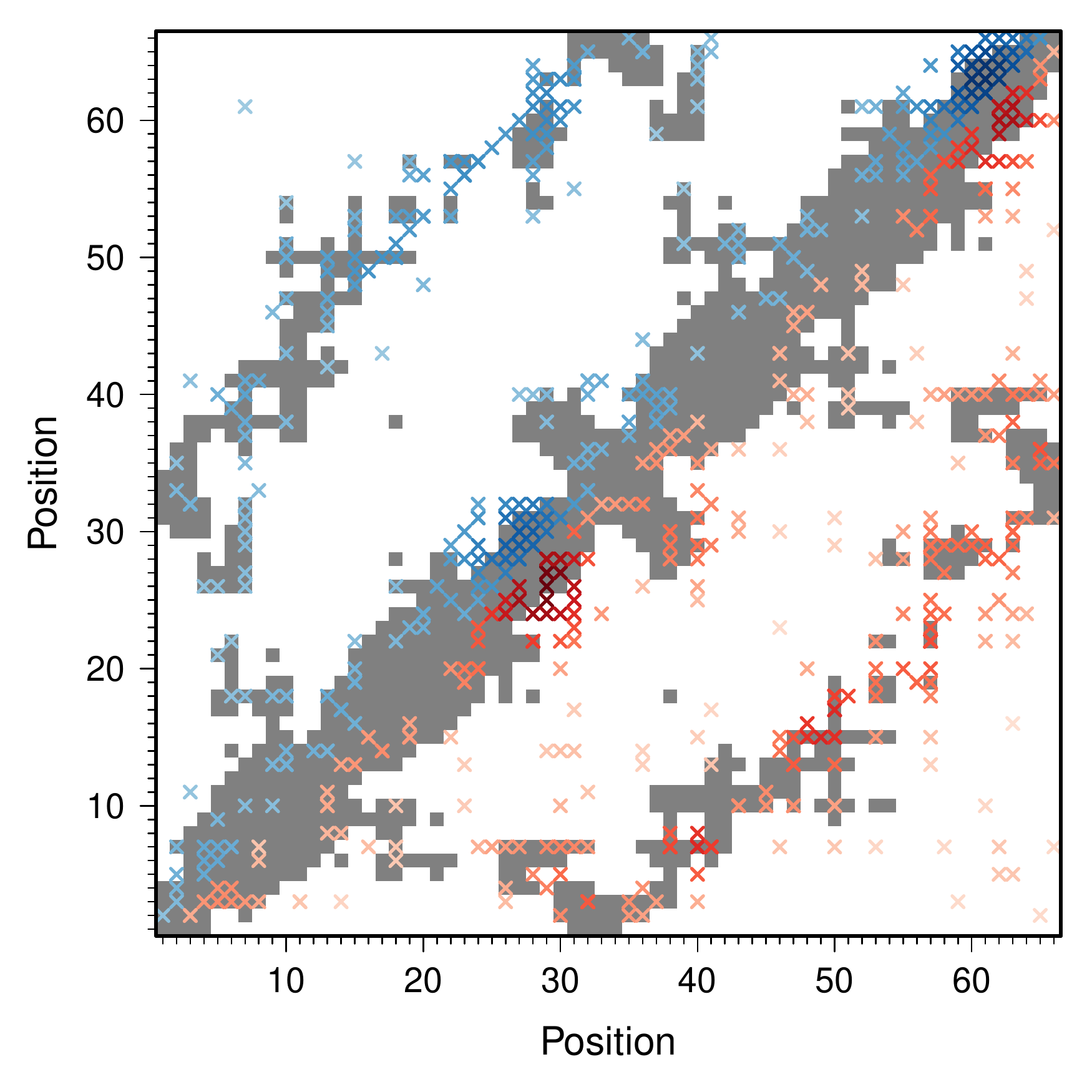}
\includegraphics[width=.3\textwidth]{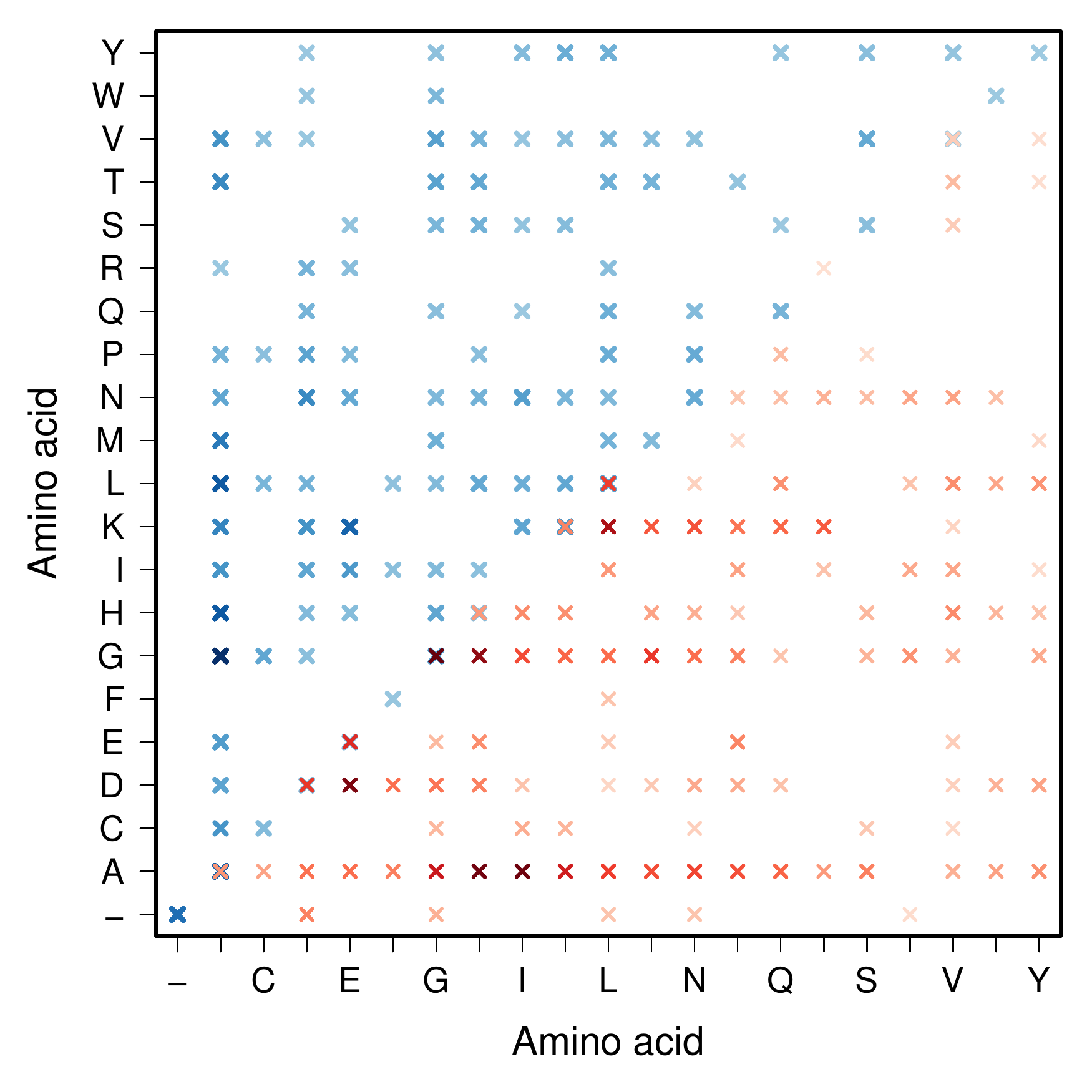}

\includegraphics[width=.3\textwidth]{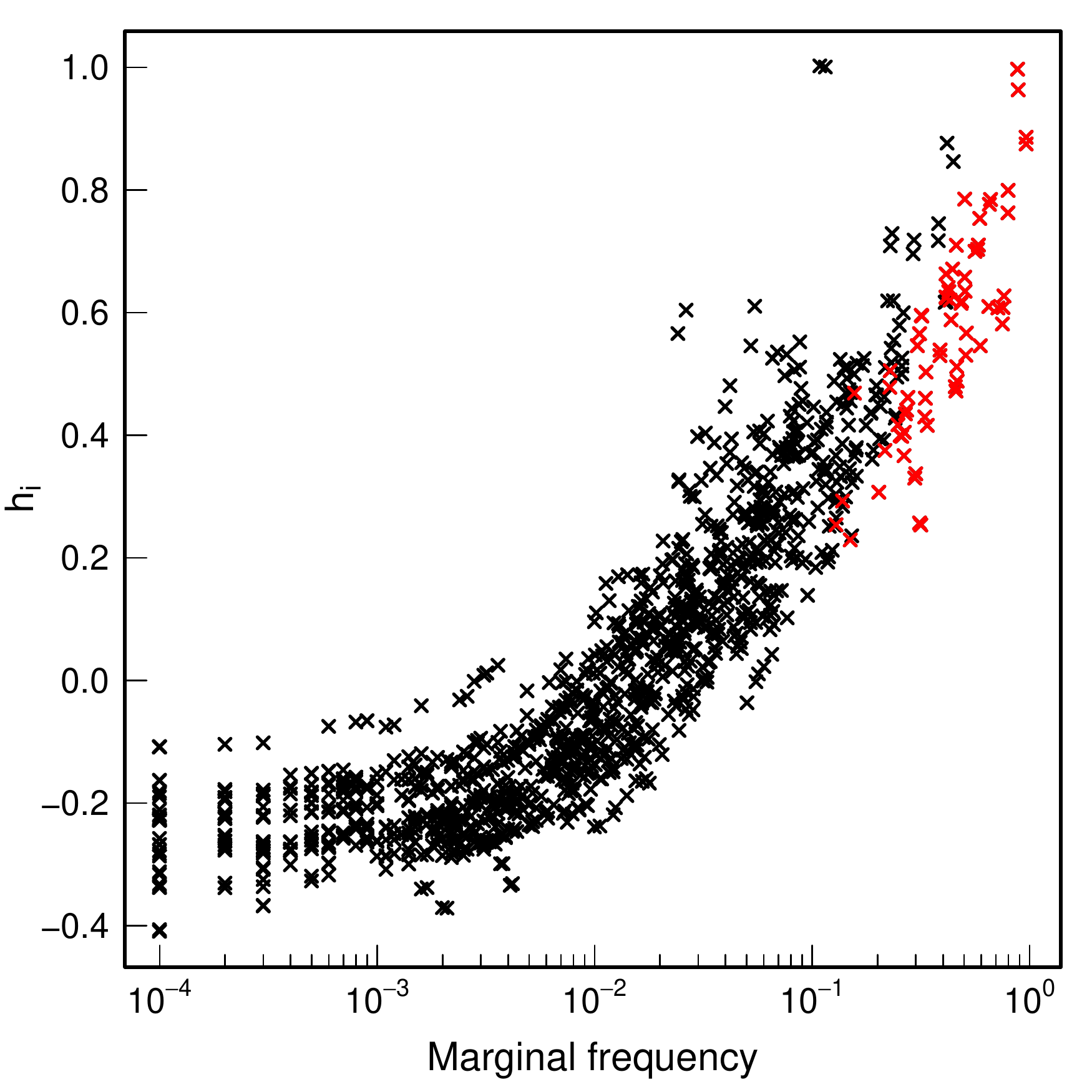}
\includegraphics[width=.3\textwidth]{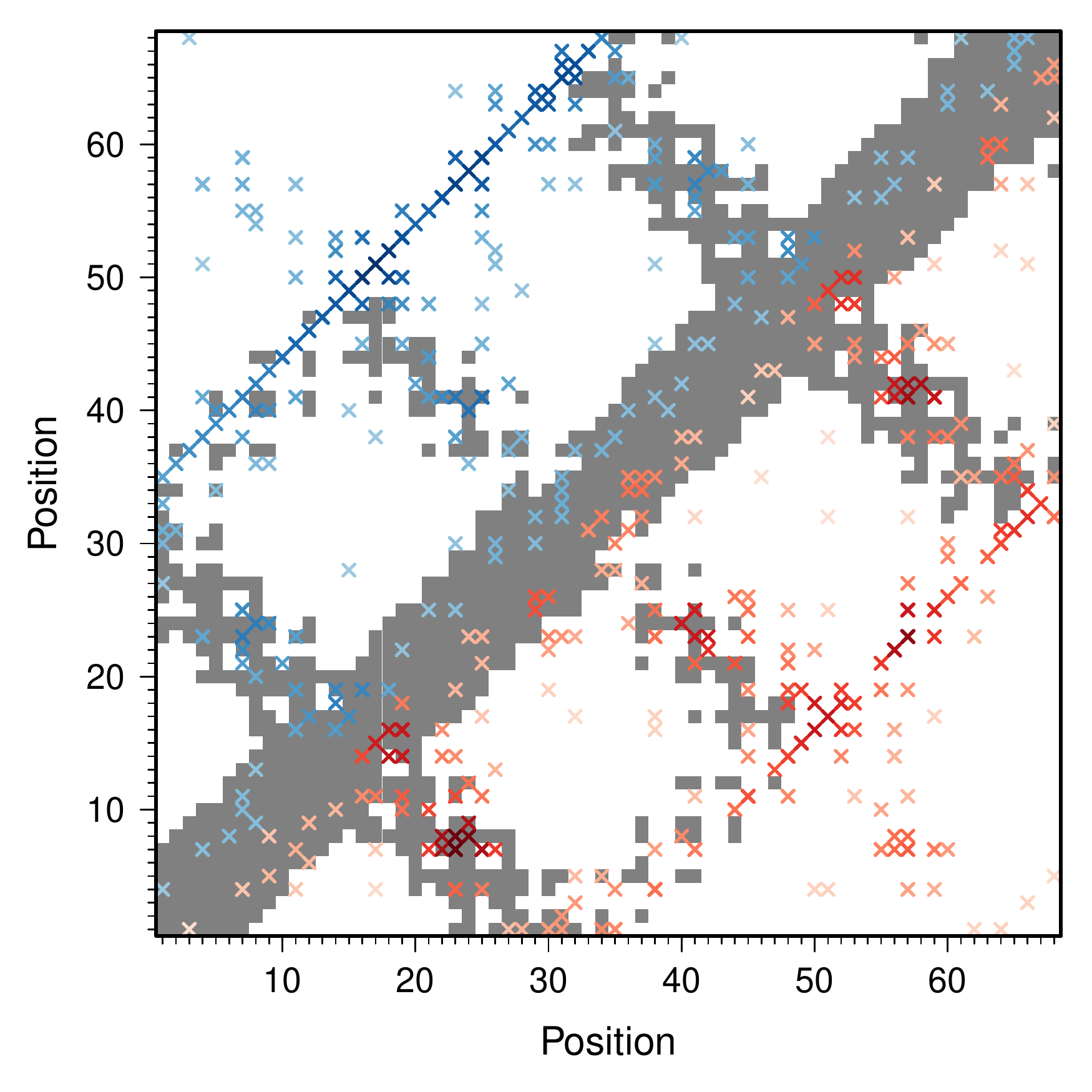}
\includegraphics[width=.3\textwidth]{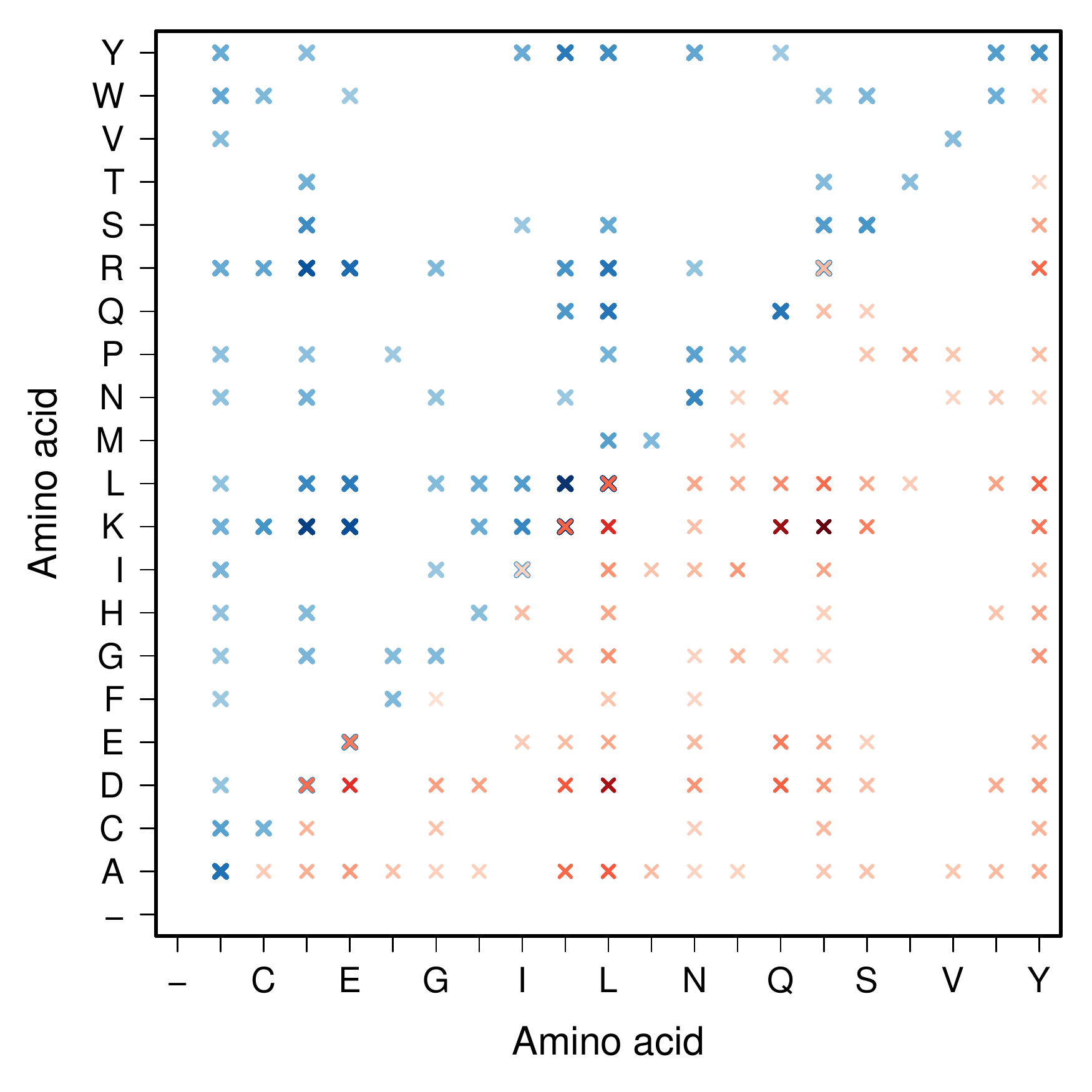}

\includegraphics[width=.3\textwidth]{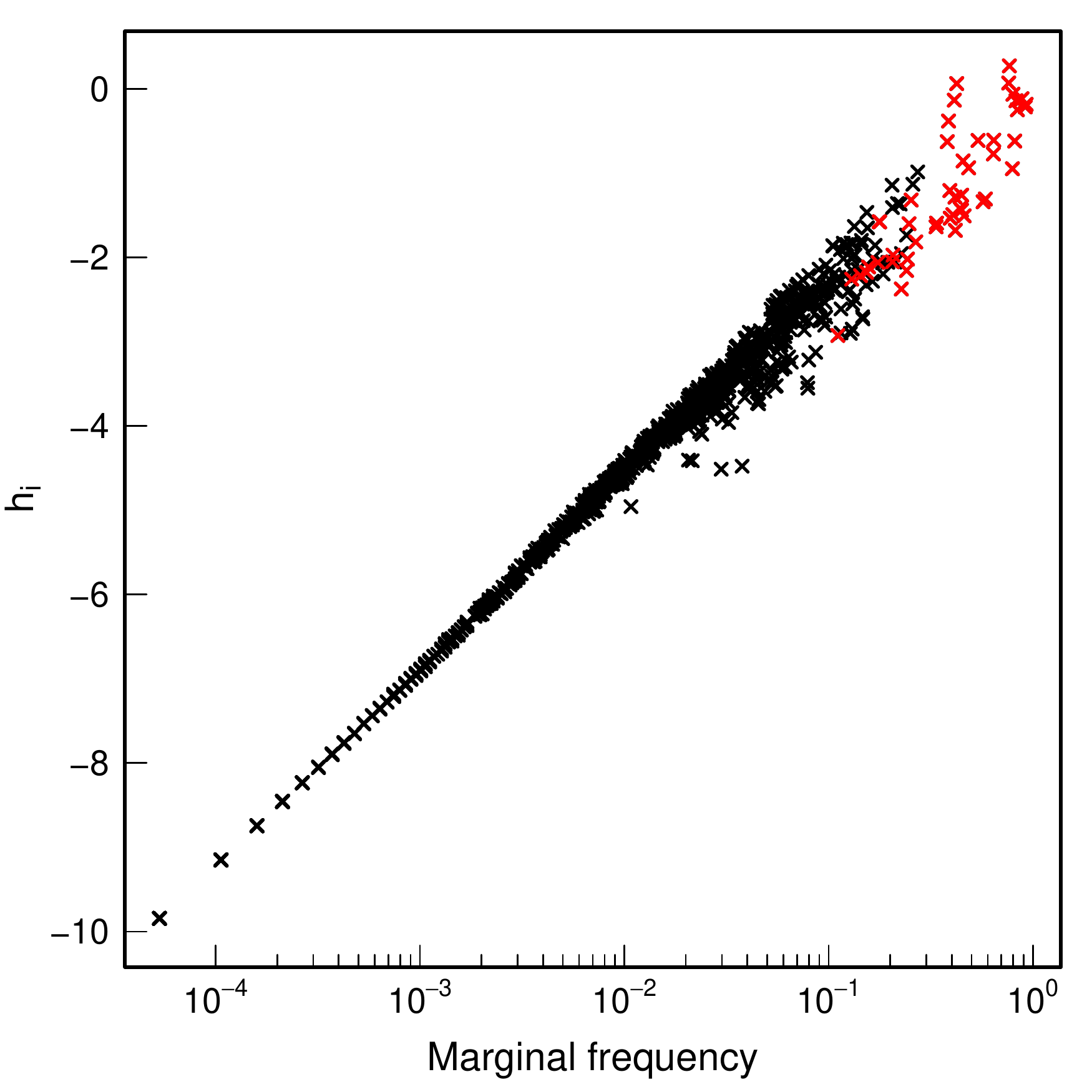}
\includegraphics[width=.3\textwidth]{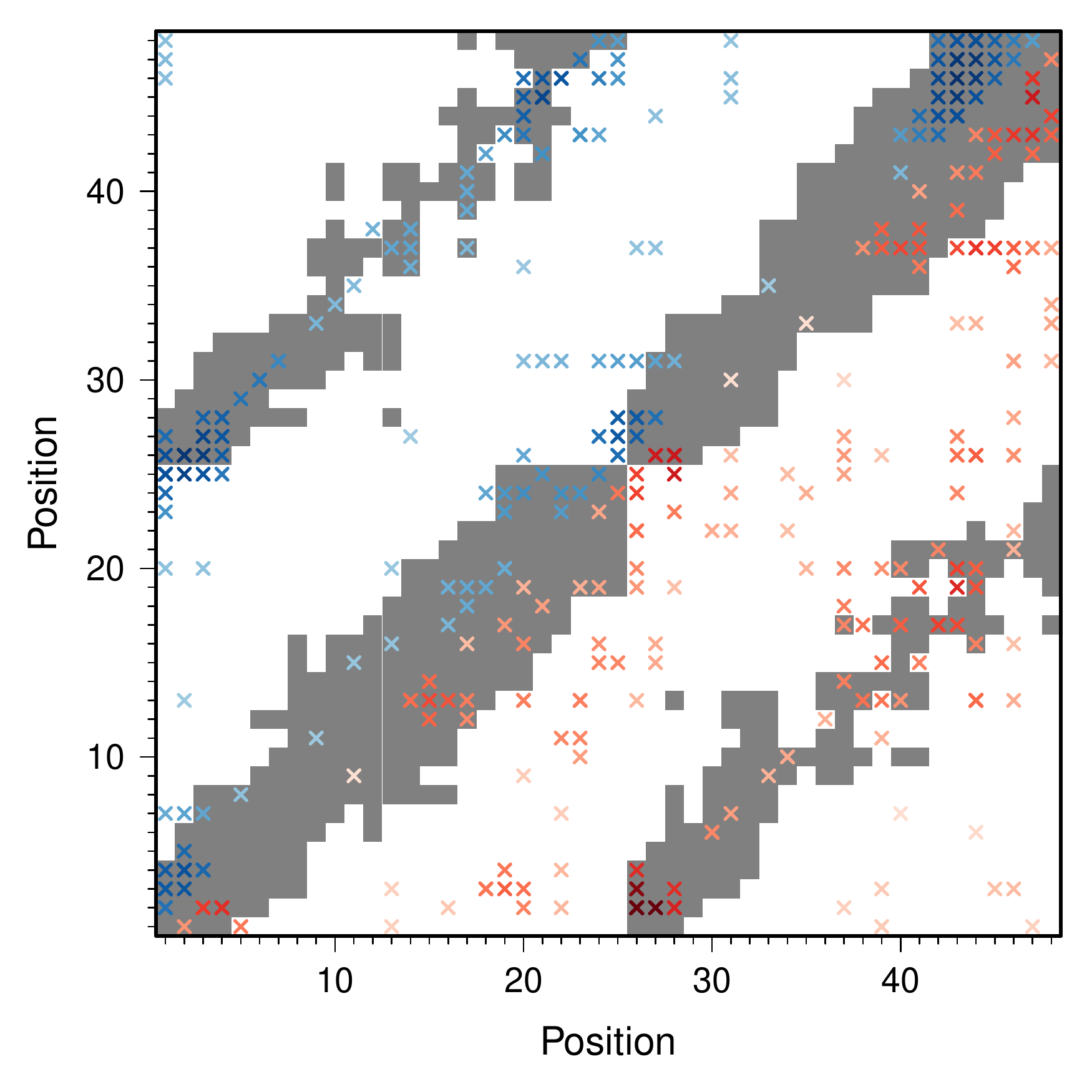}
\includegraphics[width=.3\textwidth]{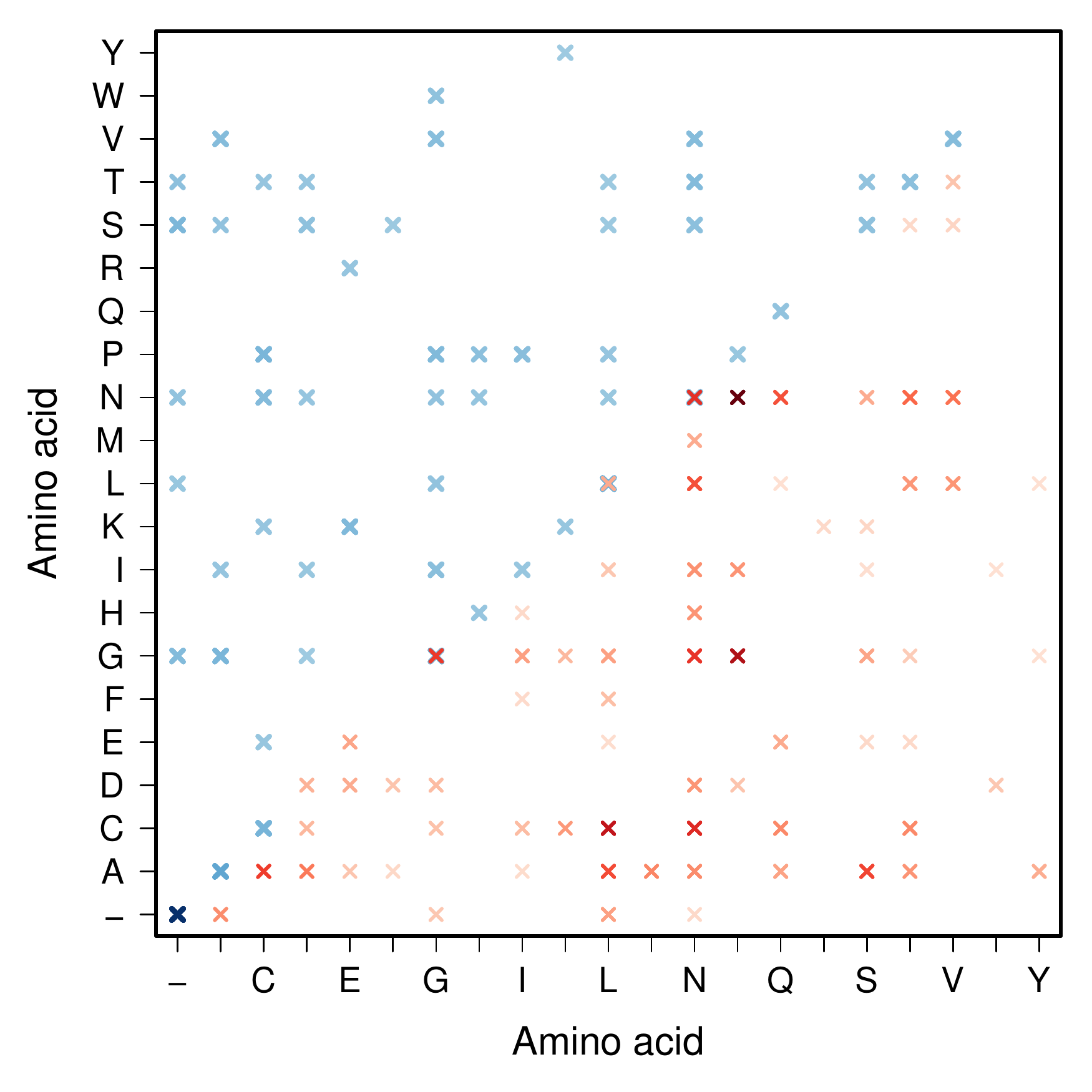}
\caption{At left, comparisson between the local field parameters $h_i(a_i)$ and the marginal frequencies $f_i(a_i)$. At center, contact map (grey indicates position in contact, white not in contact on the native structure). On blue, pairs of positions involed in highest $J_{ij}(a_i,b_j)$, red lowest $J_{ij}(a_i,b_j)$. At right, pairs of amino acids involved in the highest $J_{ij}(a_i,b_j)$ parameters (on blue) and in the lowest $J_{ij}(a_i,b_j)$ (on red). ANK at the top. TPR at the center. LRR at the bottom.}
\label{fig:campos}
\end{figurehere}
\medskip

\end{document}